\documentclass[aps,prb,reprint,superscriptaddress]{revtex4-2}

\usepackage{amsmath}
\usepackage{amssymb,amsfonts}
\usepackage{bm}

\usepackage{physics}

\usepackage{graphicx}
\usepackage{xcolor}

\usepackage{array}
\usepackage{booktabs}
\usepackage{dcolumn}
\usepackage{multirow}

\usepackage{algorithm}
\usepackage{algorithmic}

\newcommand{\MC}[1]{\textcolor{black}{{#1}}}

\begin{document}
\title{Systematically Improvable Numerical Atomic Orbital Basis Using Contracted Truncated Spherical Waves}

\author{Yike Huang}
\affiliation{Institute of Microelectronics, Chinese Academy of Sciences, Beijing 100029, China}
\affiliation{AI for Science Institute, Beijing 100080, China}

\author{Zuxin Jin}
\affiliation{AI for Science Institute, Beijing 100080, China}

\author{Linfeng Zhang}
\affiliation{AI for Science Institute, Beijing 100080, China}
\affiliation{DP Technology, Beijing 100080, China}

\author{Mohan Chen}
\email{mohanchen@pku.edu.cn}
\affiliation{AI for Science Institute, Beijing 100080, China}
\affiliation{\mbox{HEDPS, CAPT, School of Mechanics and Engineering Science and School of Physics}, Peking University, Beijing 100871, China}

\begin{abstract}
To solve the Kohn-Sham equations within the framework of density functional theory, we develop a scheme to construct numerical atomic orbital (NAO) basis sets by contracting truncated spherical waves (TSWs). 
The contraction minimizes the trace of the kinetic operator in the residual space, generalizing the spillage minimizing scheme [M. Chen \textit{et al.}, J. Phys. Condens. Matter \textbf{22}, 445501 (2010); P. Lin \textit{et al.}, Phys. Rev. B \textbf{103}, 235131 (2021)].
In addition to the systematic improvability inherited from previous schemes, the use of TSWs instead of plane waves as the expansion basis bridges reference states and NAOs more effectively, and eliminates spurious interactions between periodic images, thereby enabling better transferability through the inclusion of extensive reference states. 
Benchmarks demonstrate that the constructed NAO achieves satisfactory precision for various properties of both molecules and bulk systems, including total energy, bond length, atomization energy, lattice constant, cohesive energy, band gap, and energy-level alignment. 
Furthermore, by incorporating unoccupied states, the improved transferability in describing the conduction band is demonstrated to be effective and substantial.
\end{abstract}

\maketitle

\section{Introduction}
\MC{
Density functional theory (DFT)~\cite{HohenbergKohn1964,KohnSham1965} is now the workhorse of first-principles electronic-structure calculations across chemistry, materials science, and condensed-matter physics. Atomic orbital basis sets strike a balance between computational efficiency and accuracy. For isolated molecules, well-developed atomic orbital basis sets already reach meV-level convergence in total energy~\cite{Blum2009}.
In general, the number of atomic orbital basis sets is smaller than that of alternative options, and computational costs can be further reduced by leveraging their locality to implement linear-scaling methods~\cite{FonsecaGuerra1998, Artacho1999}.
Owing to these inherent advantages, they have been widely employed for atomic systems of various sizes, ranging from a few atoms up to thousands of atoms.~\cite{abacuslargescale2025,cp2klargescale2014,orcalargescale2020,pyscflargescale2019,nwchemlargescale2020,siestalargescale2020,fhiaimslargescale2025,openmxlargescalegpu2025,gaussian2016,shao2015qchem,werner2012molpro,barca2020gamessus,guest2005gamessuk,ahlrichs1989turbomole,turney2012psi4}.
}

\MC{
However, atomic orbital bases generally lack the systematic convergence inherent to other basis types. For instance, the completeness of the plane-wave basis can be controlled by adjusting the kinetic energy cutoff. Even among localized basis sets, there also exist systematically refinable alternatives such as real-space grids~\cite{Beck2000}, truncated spherical waves~\cite{Haynes1997}, B-splines~\cite{Hernandez1997}, Lagrange functions~\cite{Varga2004}, wavelets~\cite{Genovese2008}, etc.
%While this fact poses no fundamental difficulty, additional effort is required to generate a sensible hierarchy of basis sets with transferable accuracy.
For atomic orbitals, additional work is needed to establish a well‑defined hierarchy of basis sets with transferable accuracy.
}

\MC{
Numerical atomic orbitals (NAOs)~\cite{Sankey1989,Porezag1995,Horsfield1997, Junquera2001,Ozaki2003,Blum2009,Chen2010} have undergone extensive development over the past few decades, with numerically tabulated radial functions that are fully flexible within a cutoff radius. In fact, the strategy for generating strictly localized NAOs is not unique. 
For example, a wide range of methods involve finding the Kohn-Sham orbitals of isolated atoms under suitable confining potentials; these are commonly referred to as pseudo-atomic orbitals, or PAOs.~\cite{Sankey1989,Porezag1995,Horsfield1997, Junquera2001,Soler2002,lehtola2025}. 
Based on this idea, Ozaki~\cite{Ozaki2003,Ozaki2004} proposed using PAOs as ``primitive functions'' to construct NAOs, with coefficients optimized during self-consistent cycles. %The resulting NAOs are strongly tied to the local chemical environment and thus do not constitute conventional basis sets.
Alternatively, Blum \textit{et al.}~\cite{Blum2009} employed the occupied PAOs as a minimal basis set and expanded it by iteratively selecting the function that most improves a target energy from a predefined pool of candidates. 
%These candidates include cation-like and hydrogen-like radial functions, with the target energy exemplified by the non-self-consistent total energies of several selected dimers.
%
Furthermore, the concept of ``correlation consistency'' introduced by Dunning~\cite{dunning1989cc} has also been investigated in the context of NAOs by Zhang \textit{et al.}~\cite{Zhang2013}.
By minimizing the frozen-core RPA@PBE total energies of free atoms, the resulting basis sets are well suited for correlation methods involving explicit summations over unoccupied states, such as RPA and MP2.
}

\MC{
Methods that adopt strategies other than explicit energy optimization also exist. For instance, S{\'{a}}nchez-Portal \textit{et al}~\cite{Sanchez-Portal1995,Sanchez-Portal1996} proposed to generate NAOs towards selected reference states by optimizing the defined ``spillage'' value.
In their original scheme, the primitive functions that constitute NAOs are PAOs or Slater-type orbitals, and the reference systems are solids.
Following their work, Kenny \textit{et al}~\cite{Kenny2000} used confined neutral/charged atoms instead of solids as reference systems.
Later, Chen, Guo, and He (CGH) ~\cite{Chen2010} proposed using truncated spherical waves (TSW) as primitive functions to gain greater flexibility and a series of dimers of variable bond lengths as reference systems to improve transferability. 
Recently, Lin, Ren, and He (LRH)~\cite{Lin2021} showed that CGH basis sets can be further improved by introducing an extra gradient term to the original spillage function.
The above-developed NAOs attain an outstanding balance of efficiency and precision, facilitating accurate large-scale simulations in a wide range of research areas.~\cite{fhiaimslargescale2025,siestalargescale2020,pham2025openmxappli,soni2025siestaappli,motlak2025siestaappli,calderan2026fhiaimsappli,mandzhieva2025fhiaimsappli,perevozchikov2025siestaappli,abacuslargescale2025,liu2023siestaappli,box2023fhiaimsappli,kumar2022siestaappli,nair2025fhiaimsappli,klein2021fhiaimsappli,montes2025siestaappli,oudhia2024siestaappli,pandey2024siestaappli,stark2025fhiaimsappli,tellez2024siestaappli,ibrahim2024siestaappli,latypov2024siestaappli,biliroglu2025fhiaimsappli,frankcombe2025fhiaimsappli,li2025abacusappli,ru2025abacusappli,liu2021abacusappli,liu2022abacusappli,luo2024abacusappli,sun2025abacusappli,jing2025abacusappli,zhang2025abacuappli,zhang2025abacusappli,miyata2018openmxappli,tatetsu2018openmxappli}.
}

\MC{
Although the NAO basis sets from Ref.~\onlinecite{Lin2021} deliver remarkable accuracy in calculating structural and electronic properties for numerous molecular and bulk systems, it is still worth studying whether this type of basis set can be further improved.
For instance, the hierarchy of these NAO basis sets only extends to standard polarization orbitals, while orbitals with higher angular momentum remain insufficiently explored, which reduces the overall completeness of the basis set.
%
%The hierarchy of LRH basis sets is built upon split-valence and polarization, where polarization orbitals are restricted to have an angular momentum one larger than that of the valence orbitals.
%The lack of involving orbitals with higher angular momentum clearly hinders the completeness: in Figure. 1 of their original work, 
%
Furthermore, the energy difference between plane waves (PWs) and TSWs with limited angular momentum can reach tens of meV. Since TSWs can be systematically refined toward a complete basis set, this energy gap may be further reduced.~\cite{bennett2025accurate}.
In addition, incorporating more reference states enhances the transferability of NAOs. Meanwhile, plane waves are generally reliable for approaching the complete basis set (CBS). However, spurious interactions with periodic images can trigger artificial bonding across the vacuum. Such issues are difficult to avoid when plane waves are adopted as the expansion basis in practical calculations (Appendix \ref{appendix:spurious-interaction-across-vacuum}).}

\MC{
In this work, we propose a scheme to construct NAO basis sets by contracting TSWs, generalizing the spillage minimization scheme via minimizing the kinetic operator trace in the residual space. Our NAOs inherit systematic improvability to approach CBS, with a rigorous strategy for determining key parameters by converging diatomic energy errors against near-complete PWs. Using TSWs as the expansion basis eliminates spurious periodic image interactions, enabling effective transferability enhancement by including virtual states in spillage, which benefits conduction band calculations. Comprehensive benchmarks on molecular and bulk systems verify that the NAOs yield satisfactory precision for key properties like total energy, lattice constant and band gap, and perform well in describing electronic structures.
}

\MC{
The rest of this paper is organized as follows. 
Section 2 describes the details of our scheme for constructing the TSW and the contraction to construct NAOs. 
Section 3 presents comprehensive benchmarking results, including total energies, bond lengths, atomization energies, and energy levels of molecules; total energies, equilibrium lattice constants, cohesive energies, and band structures of bulk materials. 
We summarize our results in Section 4.}

\section{Methods}
\MC{
An atomic orbital of the form $\phi(\bm{r}) = \chi_{l\zeta}(r) {Y}_{lm}(\hat{r})$ has all its flexibility in the radial part, 
in which the $\chi_{l\zeta}(r)$ and $Y_{lm}(\hat{r})$ stand for the radial and angular parts, respectively. $l$ denotes angular momentum, $\zeta$ distinguishes different radial functions, $Y_{lm}(\hat{r})$ is the spherical harmonics and $m$ is the magnetic quantum number.
Constructing a basis set boils down to selecting a suitable parametrization for
$\chi$ and defining an appropriate optimization problem to determine these parameters.
}

\subsection{Parametrization}
\MC{
The parametrization of the radial shape typically balances several factors, including the computational efficiency of integrals, locality, and flexibility. An ideal form features strict locality and maximum flexibility within a reasonable cutoff radius while allowing for the efficient evaluation of common integrals.
}

\MC{
Over the past few decades, several numerical algorithms have been proposed~\cite{Talman1978,Sharafeddin1992,Toyoda2010,Havu2009} to perform integrations for NAOs efficiently.
For strict locality and maximum flexibility, TSWs~\cite{Haynes1997} emerge as a suitable basis for generating NAOs, which gives 
\begin{align}
    \chi_{l\zeta}(r) = \left\{
    \begin{matrix}
        \displaystyle\sum_{q=1}^{N_{l}} j_l(\theta_{lq}r/r_\mathrm{c}) c_{lq\zeta} & r \leq r_\mathrm{c} \\[6pt]
        0 & r > r_\mathrm{c}
    \end{matrix}
    \right.
~~.
\end{align}
Here $j_l$ is the spherical Bessel function of the first kind, $\theta_{lq}$ is the $q$-th positive zero of $j_l$, $r_\mathrm{c}$ is the cutoff radius, and $c_{lq\zeta}$ is the linear combination coefficient.
Note that spherical waves are also eigenfunctions of the kinetic energy operator (assuming atomic Rydberg unit)
\begin{align}
    -\nabla^2 (j_l(kr) Y_{lm}(\hat{r})) = k^2  (j_l(kr) Y_{lm}(\hat{r}))~~,
\end{align}
where $k$ is the norm of momentum.
Therefore, similar to the plane wave basis, the number of spherical waves $N_l$ can be controlled by a kinetic energy cutoff $E_\mathrm{c}$ 
\begin{align}
    N_{l}=\text{max}\{ q |\theta_{lq} < r_c\sqrt{E_\mathrm{c}} \}~~.
\end{align}
The full set of TSWs converges to the complete basis set (CBS) as the cutoff radius, maximum angular momentum, and cutoff energy all tend to infinity. For a fixed cutoff radius, constructing an atomic orbital basis is equivalent to determining the contraction coefficients of TSWs.}

\MC{
Note that TSWs do not possess vanishing derivatives at the cutoff radius. To avoid derivative discontinuities, previous works~\cite{Kenny2000,Chen2010} have employed an additional smoothing function.
Although this modification has no significant impact on the quality or implementation of NAOs, it not only requires extra effort to determine an optimal smoothing parameter in addition to optimizing the contraction coefficients, but also prevents the orbitals from being pure combinations of TSWs. As a result, the potential advantages offered by certain analytical integral expressions~\cite{Haynes1997, Monserrat2010} can no longer be exploited.
}

\MC{
Rather than modifying the functional form, we propose searching for the subspace where derivatives vanish at the cutoff radius while retaining the original form. Suppose we wish to combine spherical Bessel functions using a set of coefficients $K_{q\lambda}$ to form a new basis set whose first $M$ derivatives vanish at the cutoff $r_\mathrm{c}$. We have the following formulas
\begin{align}\label{eq:j2smooth}
    \xi_{l\lambda}(r)&=\sum_{q=1}^{N_l}j_l(\theta_{lq}r/r_\mathrm{c})K_{q\lambda}~~,\\
    \left.\frac{\mathrm{d}^{m}}{\mathrm{d}r^m}\xi_{l\lambda}(r)\right|_{r=r_c}&=0~~~~~~~~m=1,\ldots,M~~.
\end{align}
Let us define
\begin{align}\label{eq:jderiv}
    D_{mq}=\left.\frac{\mathrm{d}^{m}}{\mathrm{d}r^m}j_{l}(\theta_{lq}r/r_c)\right|_{r=r_c},
\end{align}
Eqs.~\ref{eq:j2smooth}-\ref{eq:jderiv} then yield $DK=0$, meaning that the columns of $K$ lie in the null space of $D$.
In practice, these can be chosen as the right singular vectors associated with zero singular values.
Moreover, since $j_l$ satisfies the spherical Bessel equation
\begin{align}
    r^2\dv[2]{j_l(kr)}{r} + 2r\dv{j_l(kr)}{r} + \left[k^2r^2 - l(l+1)\right]j_l(kr) = 0~~,
\end{align}
it follows that $D_{2q} = (-2/r_c)D_{1q}$.
This implies that any linear combination that suppresses the first derivative automatically eliminates the second derivative as well.
If only the first two derivatives are considered, we can construct a set of $N_l-1$ basis functions $\{\xi_{l\lambda}\}$ with vanishing first two derivatives from $N_l$ spherical Bessel functions. 
In the remainder of this paper, we refer to $\{\xi_{l\lambda}\}$ as NSW and unless stated otherwise, all orbitals are constructed from NSW with $M=2$.
}

\subsection{Optimization}
\begin{figure}
    \centering
    \includegraphics[width=1\linewidth]{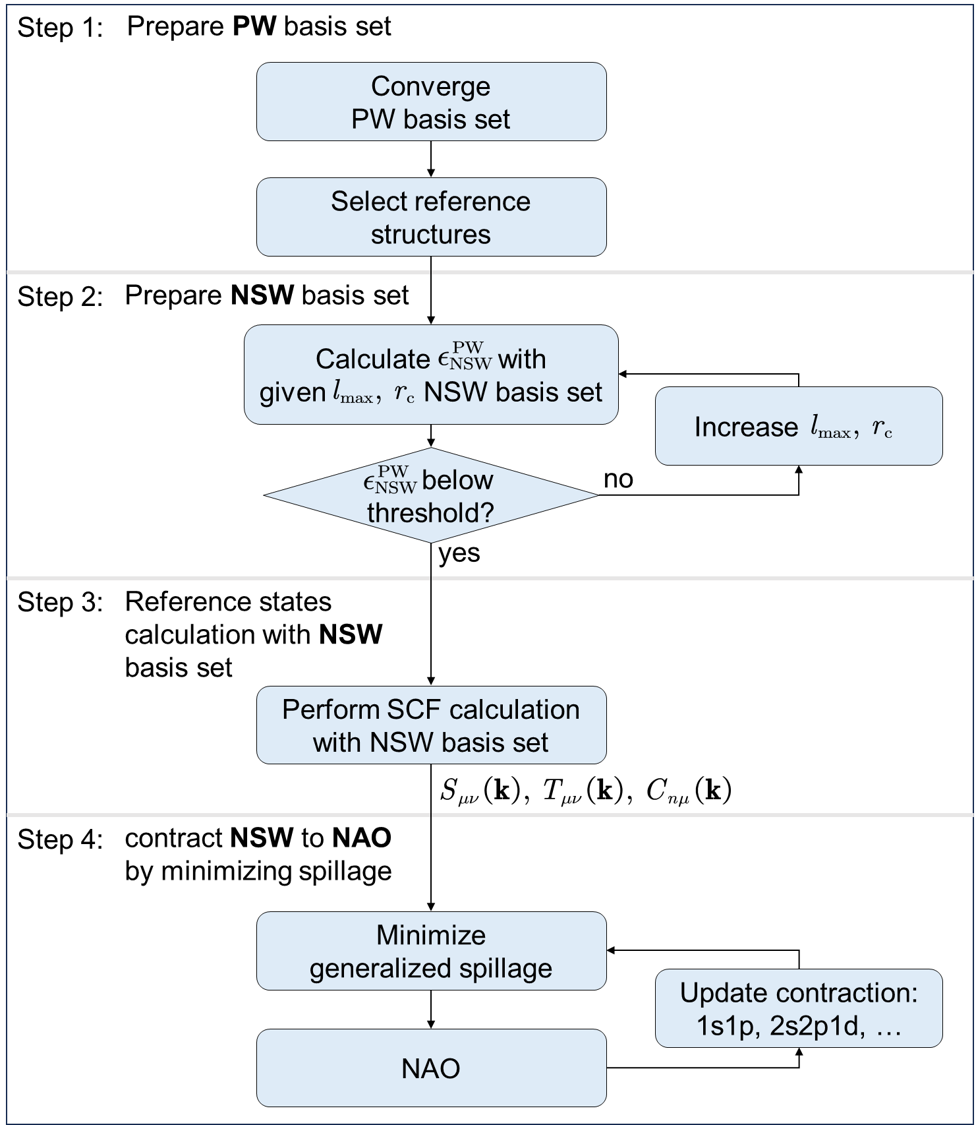}
    \caption{\MC{Workflow of NAOs generation divided into four steps. First, a plane-wave (PW) basis set is prepared and a set of reference structures is chosen. Second, the superparameters $r_\mathrm{c}$ and $l_{\max}$ are chosen such that the energy difference $\epsilon^\mathrm{PW}_\mathrm{NSW}$ between PW and NSW basis sets is below a threshold. Third, a series of SCF calculations is performed using the converged NSW basis, yield the overlap matrix $S_{\mu\nu}(\mathbf{k})$, the kinetic energy matrix $T_{\mu\nu}(\mathbf{k})$, and the wavefunction coefficients $C_{n\mu}(\mathbf{k})$, where $\mu$ and $\nu$ run over all NSW functions, $n$ denotes eigenstates, and $\mathbf{k}$ denotes different $\mathbf{k}$-points. Finally, the generalized spillage (Eq.~\ref{eq:generalized-spillage}) is minimized, and the contraction of the NSW basis is performed.}}
    \label{fig:csw-nao-workflow}
\end{figure}

\MC{
The contraction of the NSW basis set to obtain optimized NAOs is demonstrated in terms of defining the generalized spillage, choosing reference systems, and the NAO basis set hierarchy.
The summarized workflow is shown in Fig.~\ref{fig:csw-nao-workflow}.
}

\subsubsection{Generalized Spillage}
\MC{
In the CGH~\cite{Chen2010} scheme, the spillage function
\begin{align}
    \mathcal{S} &= \sum_{n\mathbf{k}} \langle \psi_{n\mathbf{k}}^\mathrm{PW}|{[1-\hat{P}_\mathbf{k}]}|\psi_{n\mathbf{k}}^\mathrm{PW}\rangle \nonumber\\
    &= \sum_{n\mathbf{k}} \left\|[1-\hat{P}_\mathbf{k}]\ket{\psi_{n\mathbf{k}}^\mathrm{PW}} \right\|^2 \label{eq:CGH's spillage}
\end{align}
is minimized, where $\{\ket{\psi_{n\mathbf{k}}^\mathrm{PW}}\}$ denotes the reference states obtained from high-quality plane-wave calculations, $n$ is the band index and $\mathbf{k}$ labels different sampling points in the Brillouin zone.
In particular, the projection operator
\begin{align}
    P({\mathbf{k}}) \equiv \sum_{\mu\nu} \ket{\phi_{\mu\mathbf{k}}} S_{\mu\nu}^{-1}(\mathbf{k}) \bra{\phi_{\nu\mathbf{k}}}
\end{align}
is defined within the $\mathbf{k}$-dependent Bloch subspace spanned by the atomic orbitals $\phi_{\mu\mathbf{k}}$, with $S_{\mu\nu}(\mathbf{k})\equiv \ip{\phi_{\mu\mathbf{k}}}{\phi_{\nu\mathbf{k}}}$ being the overlap matrix. 
Motivated by Eq.~\ref{eq:CGH's spillage} and the improved form
\begin{align}
    \mathcal{S}^\prime = \mathcal{S} + \sum_{n\mathbf{k}} \left\|\hat{p}(1-\hat{P}_\mathbf{k})\ket{\psi_{n\mathbf{k}}^\mathrm{PW}} \right\|^2\label{eq:LRH's spillage}
\end{align}
proposed by Lin \textit{et al}.~\cite{Lin2021}, in which $\hat{p}$ is the momentum operator, we introduce the function $\tilde{\mathcal{S}}$ to be minimized as
\begin{align}
    \tilde{\mathcal{S}}=\sum_{n\mathbf{k}}{\langle\psi_{n\mathbf{k}}^\mathrm{NSW}|[1-\hat{P}_\mathbf{k}]\hat{O}[1-\hat{P}_\mathbf{k}]|\psi_{n\mathbf{k}}^\mathrm{NSW}\rangle}~~.\label{eq:generalized-spillage}
\end{align}
This formulation depicts a distinct framework from those of S{\'{a}}nchez-Portal \textit{et al}., Chen \textit{et al}. (Eq.~\ref{eq:CGH's spillage}) and Lin \textit{et al}. (Eq.~\ref{eq:LRH's spillage}).
Specifically, Eq.~\ref{eq:generalized-spillage} corresponds to the trace of a general operator $\hat{O}$ evaluated in the residual subspace spanned by $\{(1-\hat{P}_\mathbf{k})|\psi_{n\mathbf{k}}^\mathrm{NSW}\rangle\}$. Here, $\{|\psi_{n\mathbf{k}}^\mathrm{NSW}\rangle\}$ denotes a set of vectors residing in the space spanned by NSWs, and the subspace spanned by the contracted NSWs (i.e., NAOs) is determined by minimizing as much as possible the residual information defined by the chosen operator.
}

\MC{
In this work, $\{|\psi_{n\mathbf{k}}^\mathrm{NSW}\rangle\}$ is chosen as the occupied bands, optionally augmented by a subset of unoccupied bands, while $\hat{O}$ is set to the kinetic operator $\hat{p}^2$.
In contrast, if $|\psi_{n\mathbf{k}}\rangle$ denotes the reference states computed with a plane-wave basis, replacing $\hat{O}$ with the overlap operator $\hat{S}$ or the sum of the overlap and kinetic operators $\hat{S}+\hat{T}$ reduces Eq.~\ref{eq:generalized-spillage} to Eq.~\ref{eq:CGH's spillage} or Eq.~\ref{eq:LRH's spillage}, respectively.
Since the space spanned by TSWs or NSWs is not strictly a subspace of plane waves due to the real-space truncation of Bessel functions at $r_\mathrm{c}$, minimizing the spillage defines a fitting-type problem.
}

\subsubsection{Reference Systems}
\MC{
To ensure the transferability of NAOs, reference systems should cover potential polarization and bonding in real chemical environments. Blum et al.~\cite{Blum2009} and Chen et al.~\cite{Chen2010} used homonuclear dimers with various bond lengths. Lin et al.~\cite{Lin2021} included additional trimers in the generation of triple-zeta level NAOs. Their NAOs all show stable precision across a wide range of systems. In this paper, we use at least 4 bond lengths for each dimer as references. The specific bond lengths are selected so that the corresponding energies relative to equilibrium fall within an energy window of approximately 1.0 eV.
}

\subsection{Basis Set Hierarchy}\label{sec:basis-hierarchy}
\MC{
To systematically approach the maximum angular momentum of NSWs, we use Dunning's hierarchy~\cite{dunning1989cc} to construct NAOs. We define three standard NAO tiers: minimal, polarized valence double-zeta (pVDZ), and polarized valence triple-zeta (pVTZ), together with two variants with constrained angular momentum, labeled pVTZ$^-$ and pVQZ$^=$, for comparison.
}

\MC{
Our method uses the “shell-wise” optimization scheme developed for the CGH~\cite{Chen2010} and LRH~\cite{Lin2021} basis sets, in which basis functions are added and optimized incrementally while keeping all previously generated orbitals fixed. The process starts by constructing a minimal basis corresponding to the valence-electron configuration of the pseudopotential.
New functions are then added and optimized with the minimal basis held fixed to form the pVDZ set.
In subsequent steps, further functions are added under fixed pVDZ constraints to produce the angular-momentum-restricted triple-zeta basis pVTZ$^-$.
Finally, further extensions yield the full triple-zeta pVTZ and the doubly restricted quadruple-zeta pVQZ$^=$ basis sets.
}

% for instance, oxygen (with valence electrons 2s and 2p, and a pseudized 1s orbital) yields a minimal basis denoted 1s1p. 
% (e.g., expanding 1s1p to 2s2p1d). 
%(e.g., 2s2p1d to 3s3p2d). 
%(e.g., 3s3p2d to 3s3p2d1f for pVTZ, and to 4s4p3d for pVQZ$^=$).
% The correspondence between the notation of NAOs and the numbers of basis functions of each angular momentum is listed in Table~\ref{tab:nao-contraction-rcut}.

\begin{table*}[]
\centering
\caption{Selection on orbital super-parameters ($r_\mathrm{c}$ in a.u.) and numbers of $\zeta$-functions for each angular momentum of elements.}
\label{tab:nao-contraction-rcut}
\begin{tabular}{@{}llllllll@{}}
\toprule
\multirow{2}{*}{Element} &
  \multirow{2}{*}{$r_\mathrm{c}$} &
  \multicolumn{5}{c}{Contraction} &
  \multicolumn{1}{c}{\multirow{2}{*}{NSW}} \\ \cmidrule(lr){3-7}
 &
   &
  \multicolumn{1}{c}{minimal} &
  \multicolumn{1}{c}{pVDZ} &
  \multicolumn{1}{c}{pVTZ$^-$} &
  \multicolumn{1}{c}{pVTZ} &
  \multicolumn{1}{c}{pVQZ$^=$} &
  \multicolumn{1}{c}{} \\ \midrule
H  & 8  & 1s     & 2s1p     & 3s2p     & 3s2p1d     & 4s3p     & 24s23p23d       \\
Li & 14 & 2s     & 4s1p     & 6s2p     & 6s2p1d     & 8s3p     & 43s43p42d       \\
B  & 10 & 1s1p   & 2s2p1d   & 3s3p2d   & 3s3p2d1f   & 4s4p3d   & 30s30p29d29f    \\
C  & 7  & 1s1p   & 2s2p1d   & 3s3p2d   & 3s3p2d1f   & 4s4p3d   & 21s20p20d19f    \\
N  & 8  & 1s1p   & 2s2p1d   & 3s3p2d   & 3s3p2d1f   & 4s4p3d   & 24s23p23d22f22g \\
O  & 7  & 1s1p   & 2s2p1d   & 3s3p2d   & 3s3p2d1f   & 4s4p3d   & 21s20p20d19f19g \\
F  & 8  & 1s1p   & 2s2p1d   & 3s3p2d   & 3s3p2d1f   & 4s4p3d   & 24s23p23d22f22g \\
Na & 14 & 2s1p   & 4s2p1d   & 6s3p2d   & 6s3p2d1f   & 8s4p3d   & 43s43p42d42f    \\
Mg & 12 & 2s1p   & 4s2p1d   & 6s3p2d   & 6s3p2d1f   & 8s4p3d   & 37s36p36d35f    \\
Al & 12 & 2s2p   & 4s4p1d   & 6s6p2d   & 6s6p2d1f   & 8s8p3d   & 37s36p36d35f    \\
Si & 11 & 1s1p   & 2s2p1d   & 3s3p2d   & 3s3p2d1f   & 4s4p3d   & 34s33p33d32f    \\
P  & 10 & 1s1p   & 2s2p1d   & 3s3p2d   & 3s3p2d1f   & 4s4p3d   & 30s30p29d29f28g \\
S  & 10 & 1s1p   & 2s2p1d   & 3s3p2d   & 3s3p2d1f   & 4s4p3d   & 30s30p29d29f28g \\
Cl & 9  & 1s1p   & 2s2p1d   & 3s3p2d   & 3s3p2d1f   & 4s4p3d   & 27s27p26d26f25g \\
Zn & 11 & 2s1p1d & 4s2p2d1f & 6s3p3d2f & 6s3p3d2f1g & 8s4p4d3f & 34s33p33d32f32g \\
Ga & 13 & 1s1p1d & 2s2p2d1f & 3s3p3d2f & 3s3p3d2f1g & 4s4p4d3f & 49s49p48d48f47g \\
As & 11 & 1s1p   & 2s2p1d   & 3s3p2d   & 3s3p2d1f   & 4s4p3d   & 34s33p33d32f32g \\
Se & 10 & 1s1p   & 2s2p1d   & 3s3p2d   & 3s3p2d1f   & 4s4p3d   & 30s30p29d29f28g \\
Br & 10 & 1s1p   & 2s2p1d   & 3s3p2d   & 3s3p2d1f   & 4s4p3d   & 30s30p29d29f28g \\
Cd & 14 & 2s1p1d & 4s2p2d1f & 6s3p3d2f & 6s3p3d2f1g & 8s4p4d3f & 43s43p42d42f41g \\
In & 14 & 1s1p1d & 2s2p2d1f & 3s3p3d2f & 3s3p3d2f1g & 4s4p4d3f & 43s43p42d42f41g \\
Sb & 12 & 1s1p1d & 2s2p2d1f & 3s3p3d2f & 3s3p3d2f1g & 4s4p4d3f & 37s36p36d35f35g \\
Te & 12 & 1s1p1d & 2s2p2d1f & 3s3p3d2f & 3s3p3d2f1g & 4s4p4d3f & 37s36p36d35f35g \\
I  & 11 & 1s1p1d & 2s2p2d1f & 3s3p3d2f & 3s3p3d2f1g & 4s4p4d3f & 34s33p33d32f32g \\ \bottomrule
\end{tabular}
\end{table*}

\section{Results and Discussion}
We first find that the truncation radius $r_\mathrm{c}$ and the included maximal angular momentum ($l_{\max}$) of NSW for elements that can converge the total energy error with respect to PW $\epsilon^\mathrm{PW}_\mathrm{NSW}$ in dimer systems.
%, because they are the most demanding system for revealing the completeness of atomic orbitals~\cite{Blum2009, Ozaki2003}. 
%
Next, we contract the NSW basis to construct NAOs by minimizing the generalized spillage defined in Eq.~\ref{eq:generalized-spillage}, and compare the prediction errors of various properties relative to PW and NSW results for molecular systems only.
All the DFT calculations were performed with ABACUS 3.8.4 (Atomic-orbital Based Ab-initio Computation at USTC) code~\cite{li2016abacus1, lin2024abacus2}. 
The Perdew-Burke-Ernzerhof (PBE) exchange-correlation functional~\cite{perdew1996generalized}  was used, and ion-core and core-electron interactions are represented using the SG15 Optimized Norm-Conserving Vanderbilt (ONCV) pseudopotential.~\cite{hamann2013optimized, schlipf2015optimization}. 
For isolated systems, enough vacuum was added in three directions ($X/Y/Z$) to avoid any overlap between occupied states of periodic images.
Dipole correction was applied to those asymmetric molecules to eliminate dipole-dipole interactions between images. 

Additionally, for dioxygen and disulfur, we performed spin-polarized Kohn–Sham DFT calculations combined with constrained DFT (CDFT), which explicitly constrains the electron number difference between spin-up and spin-down channels to 2, in order to obtain the energetically most favorable states.
For bulk systems, the $\mathbf{k}$-point sampling was done on a uniform grid (Monkhorst-Pack method~\cite{monkhorst1976special}) with spacing $2\pi\times$0.08 Bohr$^{-1}$. 
A Gaussian-type smearing on the electronic population was used with a spread of 0.015 Ry to improve the Self-Consistent Field (SCF) convergence efficiency. 
For relaxation, the Broyden--Fletcher--Goldfarb--Shanno (BFGS) and conjugate gradient (CG) algorithms were employed for the ion and ion-cell, respectively. 
The convergence thresholds of forces and stresses were set at 0.001 Ry/Bohr and 0.1 kbar.

\subsubsection{Converging truncated spherical wave parameters}
\begin{figure*}
    \centering
    \includegraphics[width=0.75\linewidth]{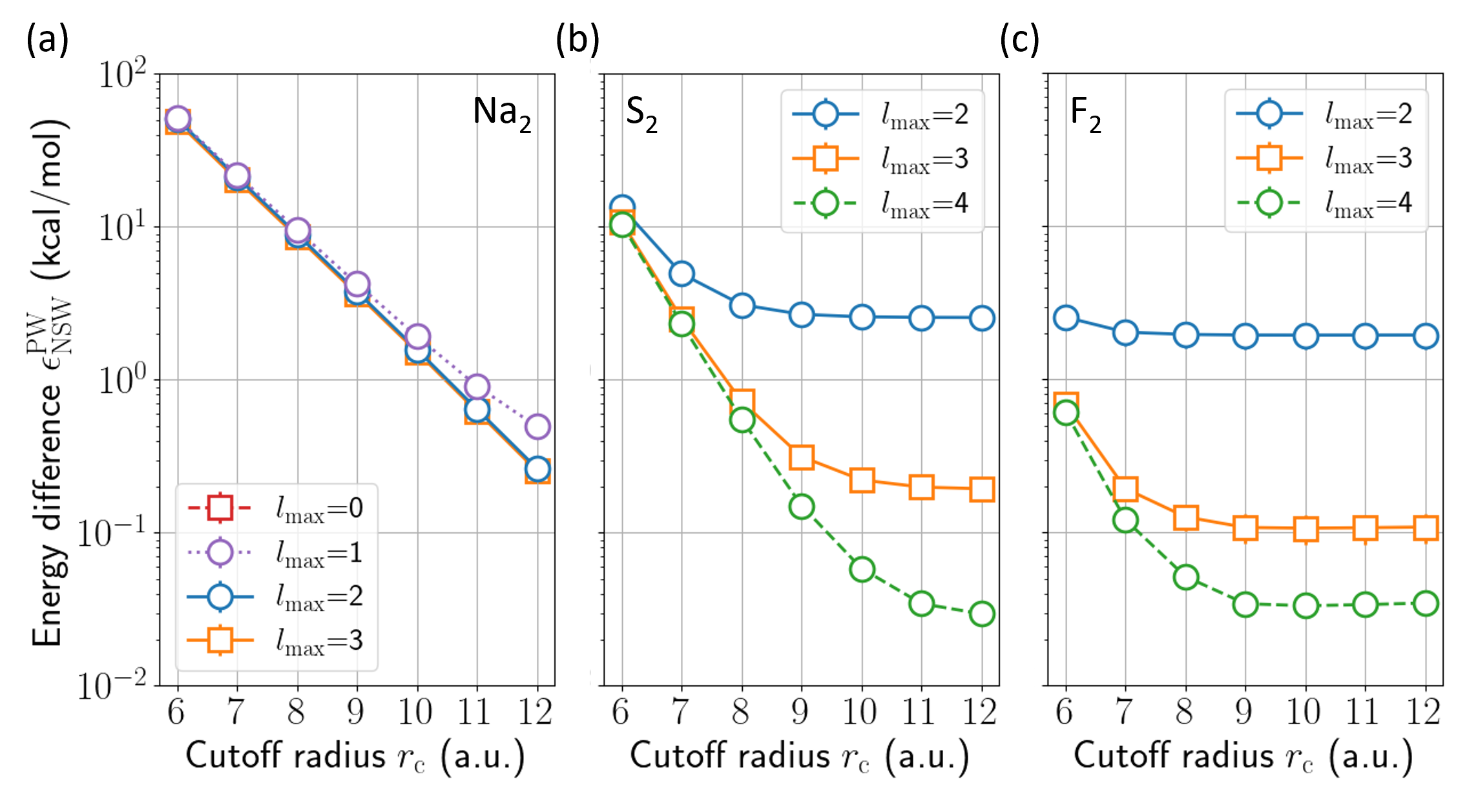}
    \caption{Systematically improbability of truncated spherical wave basis. The joint convergence of energy with respect to PW $\epsilon^\mathrm{PW}_\mathrm{NSW}(r_\mathrm{c}, l_{\max})$ on truncation cutoff radius $r_\mathrm{c}$ and maximal included angular momentum $l_{\max}$ of spherical waves tested in (a) Na, (b) S and (c) F diatomic molecule systems. The data points and error bars show the average values and standard deviations of $\epsilon^\mathrm{PW}_\mathrm{NSW}(r_\mathrm{c}, l_{\max})$ among a set of bond lengths. To distinguish between trajectories with different $l_{\max}$, lines are plotted with different markers, colors, and line styles. Red dashed with square: $l_{\max}=0$; Purple dotted with circle: $l_{\max}=1$; Blue solid with circle: $l_{\max}=2$; Orange solid with square: $l_{\max}=3$; Green dashed with circle: $l_{\max}=4$.}
    \label{fig:NSW-systematicity}
\end{figure*}

With a given kinetic energy cutoff for spherical waves, the maximum angular momentum $l_{\max}$ and a uniform real-space cutoff radius $r_\mathrm{c}$ together define a unique set of NSWs. The angular momentum components included cover the range from 0 to $l_{\max}$.
By increasing $r_\mathrm{c}$ and $l_{\max}$ of NSWs, systematic converging behaviors towards PW are observed, as shown in Fig~\ref{fig:NSW-systematicity}. 
At given $l_{\max}$, $\epsilon^\mathrm{PW}_\mathrm{NSW}$ decreases when $r_\mathrm{c}$ expands and shows converging behavior. 
Interestingly, this behavior varies substantially across elements. 
For alkali metals such as sodium (Na) shown in Fig~\ref{fig:NSW-systematicity}(a), a strong dependence on the cutoff radius $r_\mathrm{c}$ is always shown; for elements of group IV-VII such as sulfur (S) and fluorine (F) in Fig~\ref{fig:NSW-systematicity}(b) and (c), $l_{\max}$ is always more dominant than $r_\mathrm{c}$.
Only when $f$ functions corresponding to $l_{\max}$=3 or components with higher angular momentum are included can $\epsilon^\mathrm{PW}_\mathrm{NSW}$  reach the chemical accuracy level of 1 kcal/mol.
Such results are consistent with the variation in the electronegativity of elements. 
We select orbital parameters $l_{\max}$ and $r_\mathrm{c}$ as small as possible for each element that converges $\epsilon^\mathrm{PW}_\mathrm{NSW}$ to 0.1 kcal/mol (4.2 meV). 
The values of $l_{\max}$ and $r_\mathrm{c}$, and the numbers of basis functions of the NAOs and NSWs of each element are shown in Table~\ref{tab:nao-contraction-rcut}.

\subsubsection{Molecules}
\begin{figure*}
    \centering
    \includegraphics[width=1.00\linewidth]{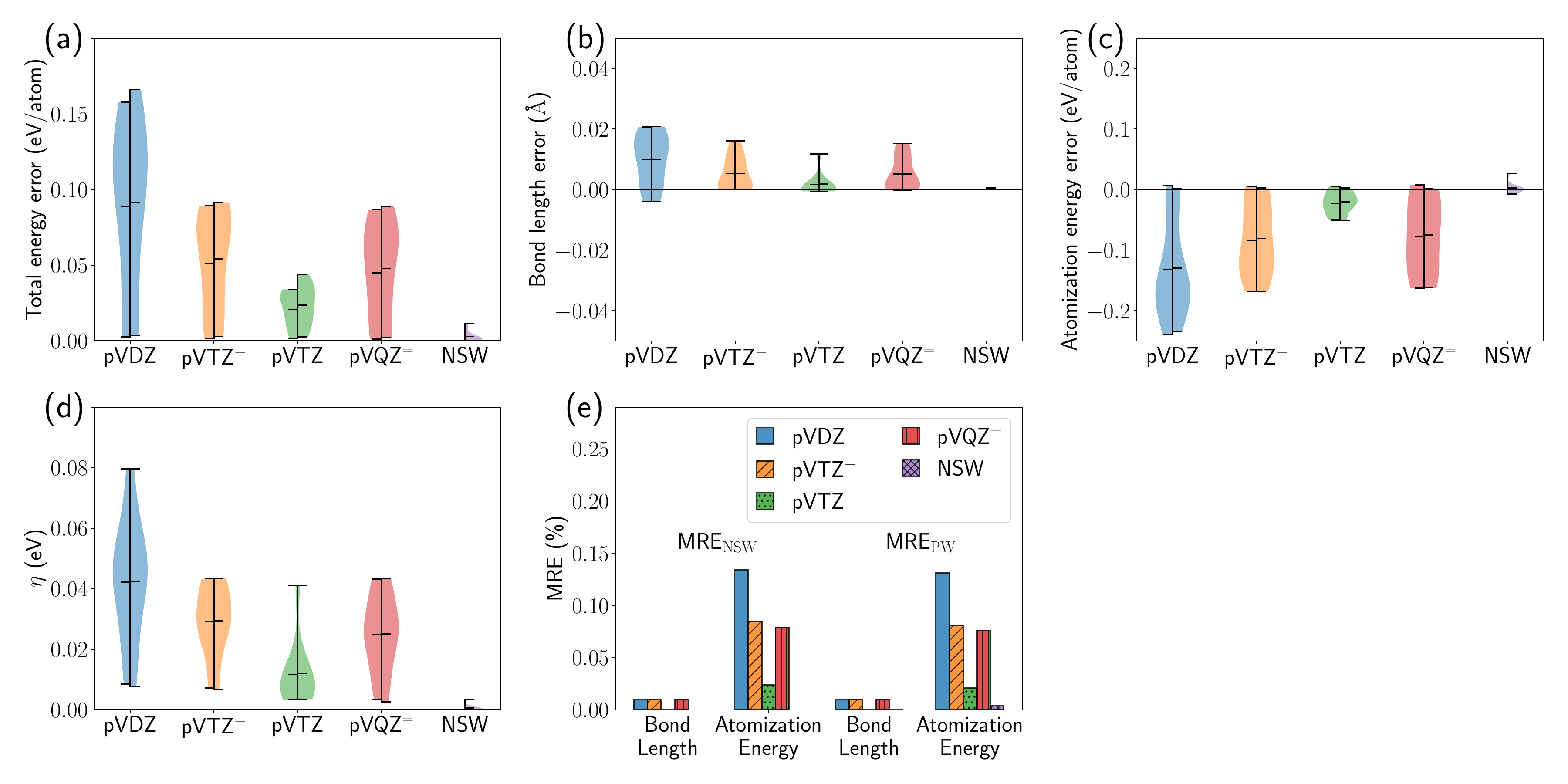}
    \caption{Molecule properties prediction benchmarks against the NSW and PW basis. For all violin plots (a-d), the left half shows the distribution of errors of NAOs with respect to the NSW basis sets, and the right half shows the distribution of errors of NAOs with respect to the PW basis sets. The extrema and the average value are indicated by black whiskers on both sides of the violin. (a) total energy error in eV/atom, which reveals the basis set completeness. (b) bond length error in \AA. (c) atomization energy error in eV/atom. (d) $\eta$-metrics in eV. (e) mean-square error with respect to the NSW and PW basis set. For figures, NAOs and NSW basis sets are distinguished by colors: blue: pVDZ; orange: pVTZ$^-$; green: pVTZ; red: pVQZ$^=$; purple: NSW. For (e), NAOs and NSW are additionally distinguished by hatches.}
    \label{fig:molecule-summary-plot}
\end{figure*}

We first performed benchmarks on the built NAOs in 11 diatomic molecular systems including $\mathrm{Br}_2$, $\mathrm{CO}$, $\mathrm{Cl}_2$, $\mathrm{F}_2$, $\mathrm{I}_2$, $\mathrm{LiH}$, $\mathrm{Li}_2$, $\mathrm{N}_2$, $\mathrm{Na}_2$, $\mathrm{O}_2$ and $\mathrm{S}_2$. 
Among these molecules, there are single ($\mathrm{LiH}$, $\mathrm{Li}_2$, $\mathrm{Na}_2$, $\mathrm{F}_2$, $\mathrm{Cl}_2$, $\mathrm{Br}_2$, $\mathrm{I}_2$), double ($\mathrm{O}_2$, $\mathrm{S}_2$) and triple ($\mathrm{CO}$, $\mathrm{N}_2$) bonds.
In the following, if not specifically mentioned, the molecules $\mathrm{O}_2$ and $\mathrm{S}_2$ are calculated for their open-shell triplet states because they possess lower energies than their singlet counterparts.
The results are summarized in Figure~\ref{fig:molecule-summary-plot}.

\paragraph{Total energies}
According to the variational theorem, the total energy serves as the primary criterion for quantitatively evaluating basis-set completeness.
We benchmarked the constructed NAOs on the 11 previously defined molecular systems to comprehensively evaluate their performance in describing wavefunctions under various chemical environments.
The PW basis set is employed to approximate the CBS, whose kinetic energy cutoff is set to the value that can converge the total energy, pressure and electron-structure well; thus, the energy differences between NAO and PW roughly indicate the error due to basis-set incompleteness relative to CBS. 
We also employ primitive basis functions (NSW) in our benchmark, with NSW serving as a reference for assessing the loss of completeness via contraction and energy differences between NSW and PW, thereby indicating the rationality of the choices of truncation radius and maximal angular momentum.

\begin{table}[]
\centering
\caption{Total energies with respect to the PW basis sets calculated with NAOs and NSW for 11 molecule systems (in eV/atom).}
\label{tab:molecule-total-energies}
\begin{tabular}{@{}llllll@{}}
\toprule
Molecule        & pVDZ   & pVTZ$^-$ & pVTZ   & pVQZ$^=$ & NSW    \\ \midrule
$\mathrm{Br}_2$  & 0.099 & 0.082 & 0.016 & 0.079 & 0.001 \\
$\mathrm{CO}$    & 0.146 & 0.074 & 0.044 & 0.057 & 0.012 \\
$\mathrm{Cl}_2$  & 0.116 & 0.092 & 0.019 & 0.089 & 0.002 \\
$\mathrm{F}_2$   & 0.086 & 0.073 & 0.035 & 0.071 & 0.001 \\
$\mathrm{I}_2$   & 0.134 & 0.038 & 0.028 & 0.032 & 0.001 \\
$\mathrm{LiH}$   & 0.028 & 0.013 & 0.012 & 0.005 & 0.000 \\
$\mathrm{Li}_2$  & 0.004 & 0.003 & 0.003 & 0.002 & 0.001 \\
$\mathrm{N}_2$   & 0.090 & 0.045 & 0.035 & 0.045 & 0.004 \\
$\mathrm{Na}_2$  & 0.004 & 0.003 & 0.003 & 0.002 & 0.001 \\
$\mathrm{O}_2$   & 0.166 & 0.087 & 0.039 & 0.065 & 0.008 \\
$\mathrm{S}_2$   & 0.135 & 0.086 & 0.026 & 0.077 & 0.001 \\
                 &       &       &       &       &       \\
MEAN             & 0.092 & 0.054 & 0.024 & 0.048 & 0.003 \\ \bottomrule
\end{tabular}
\end{table}

In Figure~\ref{fig:molecule-summary-plot}(a), the distribution of the energy error NAOs becomes narrower as the number of basis functions increases, and the violin of NSW is nearly invisible, which indicates that the real-space truncation and angular momentum selection introduce negligible error.
Table~\ref{tab:molecule-total-energies} further shows the detailed error values, and from which it can be observed that for all cases the error decreases subsequently in two basis functions increasing orders  pVDZ$\rightarrow$pVTZ$^-$$\rightarrow$pVTZ$\rightarrow$NSW, 
and pVDZ$\rightarrow$pVTZ$^-$$\rightarrow$pVQZ$^=$$\rightarrow$NSW,
among which the pVTZ can have energy error relative to PW below the chemical accuracy for all systems except the CO, where the error is only 1 meV/atom higher. 
These results suggest pVTZ has good completeness and is reliable in high-precision energy evaluation tasks.

However, basis-set completeness depends not only on the number of basis functions but also on their angular momentum.
The pVTZ and pVQZ$^=$ basis sets contain similar numbers of basis functions, according to Table~\ref{tab:nao-contraction-rcut}, but the performance differs. For example, in the $\mathrm{Br}_2$ case, the number of basis functions of pVTZ (3s3p2d1f) and pVQZ$^=$ (4s4p3d) cases are 29 and 31, respectively. In addition, the energy difference is 0.063 eV/atom, reflecting the significant difference in the ability to expand the exact wavefunction. 
This can be explained from the basis optimization based on pVTZ$^-$ (3s3p2d), because sufficient radial functions are already present to reproduce the wavefunction to a given precision; those basis functions leave the landscape of the $s/p/d$-type basis function optimization flat or highly non-convex, as no significant residual information remains to be extracted from the reference states.  
Alternatively, adding a single $f$-type basis function may be preferable, enabling further significant reduction in spillage. 
As a result, pVTZ has spillage $2.189\times10^{-4}$, while pVQZ$^=$ has $7.561\times10^{-4}$. 
Extending the range of states included in spillage (by increasing $n$ in Eq.~\ref{eq:generalized-spillage}) can also mitigate the ill-conditioning in the basis-function optimization tasks. 
More discussion will be provided in the last section and appendix.
On the other hand, the NSW basis sets serve as the primitive basis sets and exhibit the smallest energy errors, reaching as low as 2.9 meV/atom on average. 
For $\mathrm{CO}$ and $\mathrm{S}_2$ molecules, the energy errors of NSW are 12 and 1 meV/atom, respectively, which are below the jY basis in Figure 1 in LRH's work (about 30 and 7 meV/atom). 

\paragraph{Bond lengths}
Besides verifying that our NAOs can accurately describe wavefunctions, structural property prediction is among the most straightforward and meaningful molecular properties to evaluate
We benchmarked the 11 molecular systems by optimizing atomic coordinates and computing bond lengths, and compared the results with those from plane-wave basis sets (Table~\ref{tab:molecule-bond-lengths}).
Here, NSW is also used to separately evaluate the accuracy loss caused by contraction and truncation.

\begin{table*}[]
\centering
\caption{Bond lengths of 11 molecules calculated with NAOs, NSW, PW basis sets and experimental data (in \AA).}
\label{tab:molecule-bond-lengths}
\begin{tabular}{@{}llllllll@{}}
\toprule
Molecule &
  \multicolumn{1}{c}{pVDZ} &
  \multicolumn{1}{c}{pVTZ$^-$} &
  \multicolumn{1}{c}{pVTZ} &
  \multicolumn{1}{c}{pVQZ$^=$} &
  \multicolumn{1}{c}{NSW} &
  \multicolumn{1}{c}{PW} 
  & \multicolumn{1}{c}{Expt.\textsuperscript{a}} 
  \\ \midrule
$\mathrm{Br}_2$  & 2.33 & 2.33 & 2.31 & 2.33 & 2.31 & 2.31 & 2.28 \\
$\mathrm{CO}$    & 1.14 & 1.13 & 1.13 & 1.13 & 1.13 & 1.13 & 1.13 \\
$\mathrm{Cl}_2$  & 2.03 & 2.03 & 2.01 & 2.03 & 2.01 & 2.01 & 1.99 \\
$\mathrm{F}_2$   & 1.44 & 1.43 & 1.43 & 1.43 & 1.42 & 1.42 & 1.41 \\
$\mathrm{I}_2$   & 2.71 & 2.69 & 2.69 & 2.69 & 2.69 & 2.69 & 2.67 \\
$\mathrm{LiH}$   & 1.60 & 1.61 & 1.61 & 1.61 & 1.61 & 1.61 & 1.60 \\
$\mathrm{Li}_2$  & 2.73 & 2.73 & 2.73 & 2.73 & 2.73 & 2.73 & 2.67 \\
$\mathrm{N}_2$   & 1.11 & 1.10 & 1.10 & 1.10 & 1.10 & 1.10 & 1.10 \\
$\mathrm{Na}_2$  & 3.09 & 3.09 & 3.09 & 3.09 & 3.09 & 3.09 & 3.08 \\
$\mathrm{O}_2$   & 1.24 & 1.23 & 1.22 & 1.22 & 1.22 & 1.22 & 1.21 \\
$\mathrm{S}_2$   & 1.93 & 1.92 & 1.91 & 1.92 & 1.91 & 1.91 & 1.89 \\
                 &      &      &      &      &      &      & \\
MAE$_\mathrm{NSW}$ & 0.01   & 0.01   & 0.00   & 0.01   &        & & \\
MRE$_\mathrm{NSW}$ & 0.64\% & 0.32\% & 0.14\% & 0.30\% &        & & \\
MAE$_\mathrm{PW}$  & 0.01   & 0.01   & 0.00   & 0.01   & 0.00   & & \\
MRE$_\mathrm{PW}$  & 0.65\% & 0.33\% & 0.14\% & 0.31\% & 0.01\% & & \\ 
\bottomrule
\end{tabular}

\medskip
\raggedright
\footnotesize
Experimental data from: \textsuperscript{a}Huber~\cite{huber2013molblexp}.
\end{table*}

As revealed in Figure~\ref{fig:molecule-summary-plot}(b), all errors are distributed not above than 0.02 \AA, in Table~\ref{tab:molecule-bond-lengths}, all averaged absolute errors (MAE) are below 0.01 \AA.
Because the standard deviation of the bond length prediction error of the DFT functionals is about 0.06 \AA~\cite{bao2019weak}, NSWs and all our NAOs are considered as reliable, the real space truncation, angular momentum selection, and primitive functions contraction do not introduce practical error.
In addition, the difference in precision improvements between introducing 1d/1f (pVTZ) and 1s1p/1s1p1d (pVQZ$^=$) can also be observed in this test. 
pVTZ can decrease the MAE and MRE of pVTZ$^-$ by nearly one half, while the decrease by pVQZ$^=$ is nearly negligible.

\paragraph{Atomization energies}
Describing wavefunctions for both coordinated and isolated atomic states with near-quantitative precision is challenging.
This represents a key criterion in transferability tests, applicable not only to DFT functionals but also to basis sets.
Errors originating from differences in how well the exact wavefunction can be expanded between these states underlie the well-known limitation of atomic-centered orbital basis sets: the basis-set superposition error (BSSE).
Since plane-wave calculations are entirely free of BSSE, we benchmark our NAOs and NSW against reference values obtained with PW basis sets.

\begin{table*}[]
\centering
\caption{Atomization energies of 11 molecule systems calculated with NAOs, NSW, PW basis sets, and experimental data (in eV).}
\label{tab:molecule-atomization-energies}
\begin{tabular}{@{}llllllll@{}}
\toprule
Molecule &
  \multicolumn{1}{c}{pVDZ} &
  \multicolumn{1}{c}{pVTZ$^-$} &
  \multicolumn{1}{c}{pVTZ} &
  \multicolumn{1}{c}{pVQZ$^=$} &
  \multicolumn{1}{c}{NSW} &
  \multicolumn{1}{c}{PW}
  & \multicolumn{1}{c}{Expt.}
  \\ \midrule
$\mathrm{Br}_2$  & 2.419  & 2.430  & 2.561  & 2.435  & 2.586  & 2.586  & 1.97\textsuperscript{a} \\
$\mathrm{CO}$    & 11.784 & 11.881 & 11.941 & 11.909 & 11.986 & 11.959 & 11.1\textsuperscript{b} \\
$\mathrm{Cl}_2$  & 2.805  & 2.824  & 2.969  & 2.829  & 2.992  & 2.991  & 2.5\textsuperscript{b}  \\
$\mathrm{F}_2$   & 2.580  & 2.602  & 2.676  & 2.604  & 2.725  & 2.724  & 1.6\textsuperscript{b}  \\
$\mathrm{I}_2$   & 2.113  & 2.253  & 2.273  & 2.257  & 2.291  & 2.291  & 1.54\textsuperscript{a} \\
$\mathrm{LiH}$   & 2.276  & 2.295  & 2.297  & 2.308  & 2.313  & 2.310  & 2.4\textsuperscript{b}  \\
$\mathrm{Li}_2$  & 0.867  & 0.867  & 0.867  & 0.866  & 0.865  & 0.864  & 1.0\textsuperscript{b}  \\
$\mathrm{N}_2$   & 9.849  & 9.918  & 9.938  & 9.918  & 9.988  & 9.989  & 9.8\textsuperscript{b}  \\
$\mathrm{Na}_2$  & 0.765  & 0.764  & 0.764  & 0.766  & 0.759  & 0.766  & 0.7\textsuperscript{b}  \\
$\mathrm{O}_2$   & 6.629  & 6.742  & 6.836  & 6.777  & 6.868  & 6.864  & 5.1\textsuperscript{b}  \\
$\mathrm{S}_2$   & 5.163  & 5.214  & 5.335  & 5.186  & 5.335  & 5.335  & 4.4\textsuperscript{b}  \\
                 &        &        &        &        &        &        & \\
MAE$_\mathrm{NSW}$ & 0.134   & 0.085   & 0.024   & 0.079   &         & & \\
MRE$_\mathrm{NSW}$ & 3.46\%  & 2.29\%  & 0.66\%  & 2.18\%  &         & & \\
MAE$_\mathrm{PW}$  & 0.131   & 0.081   & 0.021   & 0.076   & 0.004   & & \\
MRE$_\mathrm{PW}$  & 3.38\%  & 2.22\%  & 0.59\%  & 2.06\%  & 0.14\%  & & \\ 
\bottomrule
\end{tabular}

\medskip
\raggedright
\footnotesize
Experimental data from: \textsuperscript{a}Glukhovtsev \textit{et al}.~\cite{glukhovtsev1995G2datasetBr}; \textsuperscript{b}Curtiss \textit{et al}~\cite{curtiss1991G2dataset}.
\end{table*}

In Figure~\ref{fig:molecule-summary-plot}(c) and Table~\ref{tab:molecule-atomization-energies}, it is shown that the error (NSW w.r.t. PW) due to the real space truncation and the selection on the maximal angular momentum is negligible (0.004 eV on average).
The error in the atomization energy of NAO also decreases in a manner consistent with previous benchmarks; notably, the pVTZ basis set has the smallest error among all the contracted basis sets shown and satisfies chemical accuracy. 

\paragraph{$\eta$-test}
Accurate optical spectrum calculations require precise descriptions of both occupied and virtual states across a wider energy range. 
Here we use an $\eta$ metric that is defined as~\cite{prandini2018precision}
\begin{align}
    \eta \left( A,B \right) =\min_{\omega} \sqrt{\frac{\sum_{n\mathbf{k}}{\tilde{f}_{n\mathbf{k}}\left( \varepsilon _{n\mathbf{k}}^{A}-\varepsilon _{n\mathbf{k}}^{B}+\omega \right) ^2}}{\sum_{n\mathbf{k}}{\tilde{f}_{n\mathbf{k}}}}}~~.\label{eqn:eta}
\end{align}
This metric can be understood as the occupation $f_{n\mathbf{k}}$-weighted Frobenius norm of state-wise eigenvalue $\varepsilon _{n\mathbf{k}}$ errors between two methods ($A$ and $B$), which is capable of evaluating the electronic structure prediction consistency. 
In eqn.~\ref{eqn:eta}, the minimization of $\eta$ by varying the level shift $\omega$ ensures the maximal alignment between two band structures, and the geometric average of occupation $\tilde{f}_{n\mathbf{k}}$ is defined by
\begin{align}
    \tilde{f}_{n\mathbf{k}}=\sqrt{f_{n\mathbf{k}}\left( \varepsilon _{\mathrm{f}}^{A},\sigma \right) f_{n\mathbf{k}}\left( \varepsilon _{\mathrm{f}}^{B},\sigma \right)}~~,\label{eqn:geometric-average-occ}
\end{align}
in which $f_{n\mathbf{k}}\left( \varepsilon _{\mathrm{f}}^{A},\sigma \right)$ and $f_{n\mathbf{k}}\left( \varepsilon _{\mathrm{f}}^{B},\sigma \right)$ are occupation numbers evaluated under a given smeared distribution (Gaussian smearing is used throughout this paper), $\varepsilon_\mathrm{f}$ and $\sigma$ denote the Fermi level and smearing width, respectively. 
In this test, $\eta$ is evaluated between NAOs, NSW, and PW, to quantify the precision loss from contraction, real-space truncation, and angular momentum selection. 
Superscripts are added to distinguish between $\eta$ values with respect to the PW and NSW basis sets ($\eta^\mathrm{PW}$, $\eta^\mathrm{NSW}$). 

In Figure~\ref{fig:molecule-summary-plot}(d), the violin of NSW is invisible. 
Quantitatively, in Table~\ref{tab:molecule-eta}, the $\eta$ of NSW basis sets is only 0.7 meV on average, which indicates that the NSW can predict the electronic structure nearly identical to PW.
Among all molecules, NSW can predict the energy level of LiH with an error of less than 0.1 meV. The molecule with the highest $\eta$ is CO, with a value of 3.3 meV, but it remains negligible.
For NAOs, the $\eta$ also shows a consistent decrease trend with respect to the increase of basis functions. 
Interestingly, although the errors of basis sets are almost negligible overall, and pVTZ is always the most out-performing NAOs except for $\mathrm{Li}_2$, $\mathrm{LiH}$ and $\mathrm{Na}_2$ molecules, where the pVQZ$^=$ basis sets have $\eta$ values only about half of the pVTZ basis sets.
A closer examination of the difference of radial function profiles reveals that, pVQZ$^=$ always improves the pVTZ$^-$ basis set on the description of those more delocalized states, except for Li, where the newly added 7- and 8-th s-type radial functions are highly localized. 
Note that the pseudopotential of Li used in this work is of the semi-core type, where the 1s electrons are also considered as the ``valence" electrons for transferability concerns. 
The improvement, then, can be rationalized as an improvement on the description of the 1s electrons.

According to the results shown above, the NSW can always yield negligible error relative to PW, indicating the high reliability of our truncation radius and angular momentum selection. 
Although pVTZ processes quite near number of basis functions with pVQZ$^=$, it has the most superior performance among NAOs tested, which proves our arguments on the importance of including basis functions with higher angular momentum.

\subsubsection{Bulks}

\begin{figure*}
    \centering
    \includegraphics[width=1.0\linewidth]{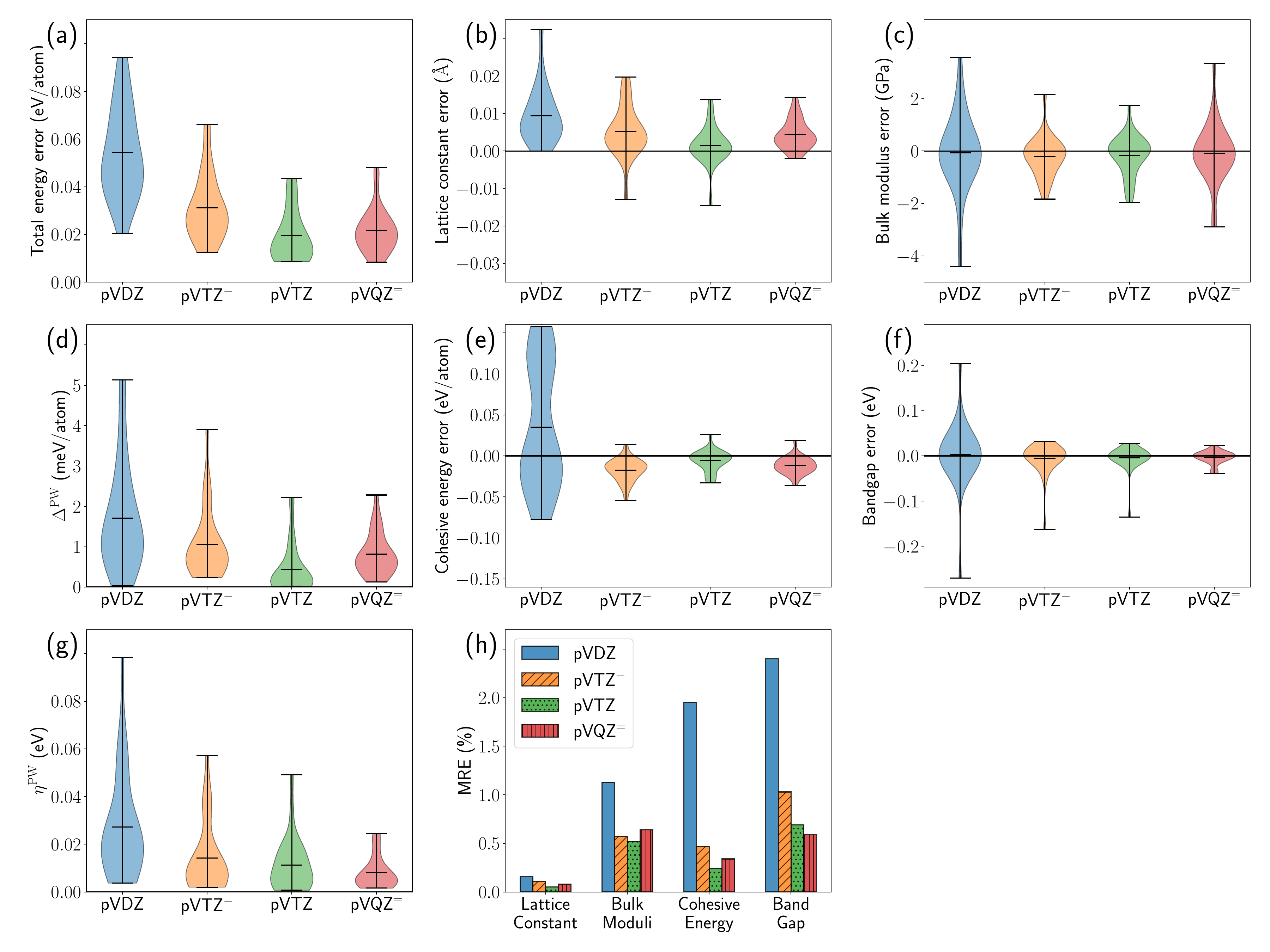}
    \caption{Bulk property prediction accuracy benchmark against the PW basis. The distribution of NAO error with respect to the PW basis sets is shown in violin plots (a-g), in which the extrema and the mean error are indicated by black whiskers. (a) total energy error in eV/atom, which reveals the basis set completeness. (b) lattice constant error in \AA. (c) bulk modulus error in GPa. (d) $\Delta$-metrics in meV/atom. (e) cohesive energy error in eV/atom. (f) bandgaps error in eV. (g) $\eta$-metrics in eV. (h) mean relative error statistics for properties including lattice constant, bulk moduli, cohesive energy, and bandgap. For figures (a-g), NAOs are distinguished by positions of violins and colors; for Figure (h), NAOs are distinguished by hatches and colors. Colors: blue: pVDZ; orange: pVTZ$^-$; green: pVTZ; red: pVQZ$^=$.}
    \label{fig:crystal-summary}
\end{figure*}

We also benchmark our NAOs in 26 bulk systems, including the classical covalent and ionic crystals, insulators and semiconductors, on the relative energies, structural (lattice constant), thermodynamic (cohesive energies), mechanical (bulk moduli) properties, and properties that are indirectly related to electron transport, including band gap and band structure relative to PW. 
Specifically, a suffix ``W" is added for ZnO wurtzite to distinguish it from the zincblende phases.
The summarized results are shown in Figure~\ref{fig:crystal-summary}.

\paragraph{Total energies}
For bulk systems, total energy serves similarly as an indicator of basis set completeness. 
Because atomic orbitals overlap more adequately in bulk, the positive energy error from incompleteness is expected to be less significant than in isolated systems. 
Here, we evaluate the completeness of our NAOs by comparing the total energies calculated with those of PW basis sets.

\begin{table}[]
\centering
\caption{Total energies with respect to the PW basis sets calculated with NAOs for 26 bulk systems (in eV/atom).}
\label{tab:crystal-total-energies}
\begin{tabular}{@{}lllll@{}}
\toprule
Bulk & 
\multicolumn{1}{c}{pVDZ} & 
\multicolumn{1}{c}{pVTZ$^-$} & 
\multicolumn{1}{c}{pVTZ} & 
\multicolumn{1}{c}{pVQZ$^=$} 
\\ \midrule
$\mathrm{AlAs}$ & 0.036 & 0.028 & 0.016 & 0.020 \\
$\mathrm{AlP}$  & 0.043 & 0.030 & 0.017 & 0.022 \\
$\mathrm{AlSb}$ & 0.031 & 0.021 & 0.015 & 0.014 \\
$\mathrm{AlN}$  & 0.093 & 0.063 & 0.034 & 0.043 \\
$\mathrm{BN}$   & 0.041 & 0.026 & 0.010 & 0.020 \\
$\mathrm{BP}$   & 0.049 & 0.028 & 0.009 & 0.022 \\
C               & 0.047 & 0.031 & 0.018 & 0.024 \\
$\mathrm{CdTe}$ & 0.066 & 0.022 & 0.019 & 0.015 \\
$\mathrm{CdS}$  & 0.056 & 0.027 & 0.018 & 0.020 \\
$\mathrm{CdSe}$ & 0.056 & 0.029 & 0.018 & 0.022 \\
$\mathrm{GaAs}$ & 0.039 & 0.017 & 0.009 & 0.013 \\
$\mathrm{GaP}$  & 0.044 & 0.018 & 0.009 & 0.015 \\
$\mathrm{GaSb}$ & 0.033 & 0.012 & 0.009 & 0.009 \\
$\mathrm{GaN}$  & 0.052 & 0.019 & 0.010 & 0.015 \\
$\mathrm{InP}$  & 0.034 & 0.019 & 0.011 & 0.016 \\
$\mathrm{LiF}$  & 0.079 & 0.047 & 0.040 & 0.014 \\
$\mathrm{MgO}$  & 0.055 & 0.034 & 0.016 & 0.025 \\
$\mathrm{MgS}$  & 0.048 & 0.032 & 0.012 & 0.023 \\
$\mathrm{NaCl}$ & 0.020 & 0.014 & 0.012 & 0.008 \\
Si              & 0.043 & 0.026 & 0.009 & 0.021 \\
$\mathrm{SiC}$  & 0.094 & 0.066 & 0.025 & 0.048 \\
$\mathrm{ZnSe}$ & 0.064 & 0.038 & 0.027 & 0.028 \\
$\mathrm{ZnTe}$ & 0.068 & 0.027 & 0.025 & 0.019 \\
$\mathrm{ZnO}$  & 0.078 & 0.047 & 0.043 & 0.029 \\
$\mathrm{ZnO}$w & 0.079 & 0.047 & 0.043 & 0.029 \\
$\mathrm{ZnS}$  & 0.069 & 0.042 & 0.032 & 0.030 \\
                &       &       &       &       \\
MEAN            & 0.054 & 0.031 & 0.019 & 0.022 \\ \bottomrule
\end{tabular}
\end{table}

In Figure~\ref{fig:crystal-summary}(a), all NAOs have smaller error averages and extrema compared with the molecular cases. 
In Table~\ref{tab:crystal-total-energies}, pVDZ has an averaged energy error with respect to the PW as 0.054 eV/atom, and there are many cases where the pVDZ can have an error satisfying the chemical accuracy ($<$ 0.042 eV/atom).
The NAO with the lowest mean energy error and narrowest error distribution is still the pVTZ basis, whose average error of 0.019 eV/atom is lower than that in the molecule system (0.024 eV/atom); all its error values are below the threshold of chemical accuracy.
In addition, the difference between the pVTZ and pVQZ$^=$ is also smaller than that in molecule systems, which suggests that the completeness improvements from higher angular momentum basis functions become less significant. In the LiF, NaCl, ZnSe, ZnTe, ZnO, ZnO2, and ZnS cases, the pVQZ$^=$ even has smaller energy errors.

\paragraph{Lattice constant}
The lattice constant is a key factor for verifying the rationality of DFT calculations. 
Although it is as crucial as bond length in isolated systems, it is more readily measurable via developed crystalline characterization techniques. 
Here we benchmark the precision of predicting lattice constants for 20 face-centered cubic crystal systems using our built NAOs, and compare the results with PW basis calculations.

\begin{table*}[]
\centering
\caption{Lattice constants of 20 face-centered cubic crystal systems calculated with NAOs, PW basis sets, and experimental data (in \AA).}
\label{tab:crystal-lattice-constant}
\begin{tabular}{@{}lllllll@{}}
\toprule
Bulk & 
\multicolumn{1}{c}{pVDZ} & 
\multicolumn{1}{c}{pVTZ$^-$} & 
\multicolumn{1}{c}{pVTZ} & 
\multicolumn{1}{c}{pVQZ$^=$} & 
\multicolumn{1}{c}{PW} 
& \multicolumn{1}{c}{Expt.}
\\ \midrule
AlAs & 5.74 & 5.73 & 5.73 & 5.73 & 5.73 & 5.66\textsuperscript{a}, 5.62\textsuperscript{b} \\
AlP  & 5.51 & 5.50 & 5.50 & 5.51 & 5.50 & 5.47\textsuperscript{a}, 5.451\textsuperscript{b} \\
AlSb & 6.23 & 6.23 & 6.22 & 6.22 & 6.22 & 6.13\textsuperscript{a}, 6.1347\textsuperscript{b} \\
BN   & 3.62 & 3.62 & 3.62 & 3.62 & 3.62 & 3.615\textsuperscript{c,b} \\
BP   & 4.55 & 4.54 & 4.54 & 4.54 & 4.54 & 4.538\textsuperscript{c,b} \\
C    & 3.56 & 3.56 & 3.56 & 3.56 & 3.56 & 3.57\textsuperscript{d}, 3.56\textsuperscript{b} \\
CdTe & 6.63 & 6.63 & 6.62 & 6.63 & 6.62 & 6.482\textsuperscript{e}, 6.480\textsuperscript{b}\\
GaAs & 5.76 & 5.75 & 5.75 & 5.75 & 5.75 & 5.65\textsuperscript{a}, 5.6537\textsuperscript{b} \\
GaP  & 5.51 & 5.51 & 5.50 & 5.51 & 5.50 & 5.45\textsuperscript{a}, 5.4505\textsuperscript{b} \\
GaSb & 6.22 & 6.21 & 6.21 & 6.22 & 6.21 & 6.10\textsuperscript{a}, 6.118\textsuperscript{b} \\
InP  & 5.97 & 5.97 & 5.96 & 5.97 & 5.96 & 5.87\textsuperscript{a}, 5.8687\textsuperscript{b} \\
LiF  & 4.06 & 4.04 & 4.04 & 4.06 & 4.06 & 4.0218\textsuperscript{f}, \\
MgO  & 4.26 & 4.25 & 4.25 & 4.25 & 4.25 & 4.2112\textsuperscript{g,b}\\
MgS  & 5.23 & 5.23 & 5.23 & 5.23 & 5.22 & 5.203\textsuperscript{b}, 5.201\textsuperscript{h}\\
NaCl & 5.70 & 5.70 & 5.70 & 5.70 & 5.70 & 5.62779, 5.63978, 5.64056\textsuperscript{b}\\
Si   & 5.48 & 5.47 & 5.47 & 5.47 & 5.47 & 5.42\textsuperscript{d}, 5.43\textsuperscript{b}\\
SiC  & 4.39 & 4.39 & 4.38 & 4.38 & 4.38 & 4.34--4.39\textsuperscript{i}\\
ZnSe & 5.74 & 5.74 & 5.73 & 5.74 & 5.73 & 5.6676\textsuperscript{b}\\
ZnTe & 6.19 & 6.19 & 6.18 & 6.19 & 6.18 & 6.089\textsuperscript{b}\\
ZnO  & 3.27 & 3.27 & 3.27 & 3.27 & 3.26 & 3.24950\textsuperscript{b} \\
ZnS  & 5.46 & 5.46 & 5.45 & 5.45 & 5.44 & 5.4093\textsuperscript{b}\\
     &      &      &      &      &      & \\
MAE$_\mathrm{PW}$ & 0.01   & 0.01   & 0.00   & 0.00   & & \\
MRE$_\mathrm{PW}$ & 0.16\% & 0.11\% & 0.05\% & 0.08\% & & \\ 
\bottomrule
\end{tabular}

\medskip
\raggedright
\footnotesize
Experimental data from: \textsuperscript{a}Vurgaftman \textit{et al}.~\cite{vurgaftman2001band}; \textsuperscript{b}Wyckoff~\cite{wyckoff1966crystal}; \textsuperscript{c}Wentzcovitch \textit{et al}.~\cite{wentzcovitch1986BNBPdataset}; \textsuperscript{d}Kittel~\cite{kittel1986introduction}; \textsuperscript{e}Horning and Staudenmann~\cite{horning1987cdte}; \textsuperscript{f}Liu \textit{et al}.~\cite{liu2007LiFbulkmoduli1}; \textsuperscript{g}Karki \textit{et al}.~\cite{karki1997MgObulkmoduli1}; \textsuperscript{h}Peiris \textit{et al}.~\cite{peiris1994MgSbulkmoduli}; \textsuperscript{i}Wang \textit{et al}.~\cite{wang2016SiCbulkmoduli}.
\end{table*}

As shown in Figure~\ref{fig:crystal-summary}(h) and Table~\ref{tab:crystal-lattice-constant}, all NAOs have the MRE with respect to the PW below 0.2\%, which is comparable to the lower bound of the standard deviation of experimental measurements.
For the pVTZ and pVQZ$^=$ basis sets, the MRE even decreases to 0.05\% and 0.08\%, respectively, indicating excellent precision.
However, although the MRE of pVTZ and pVQZ$^=$ are similar, in Figure~\ref{fig:crystal-summary}(b), it can be found that the distribution of lattice constant error of pVTZ and pVQZ$^=$ is different.
The error of pVTZ is symmetrically distributed around 0, whereas for the rest of the basis, including pVDZ, pVTZ$^-$, and pVQZ$^=$, most cases show a positive deviation.
Therefore, including basis functions with high angular momentum would be more efficient for reducing the error and is shown to be more effective in strengthening the interatomic bonds in the present cases.

\paragraph{Bulk modulus}
In addition to static structural properties such as the lattice constant, a key mechanical property is the bulk modulus, which serves as the central quantity in solid equation-of-state (EOS) fitting and is closely related to the trace of the full stress tensor. 
Bulk modulus quantifies the rigidity to isotropic inflation-compression, defining structural behavior and mechanical response near volume equilibrium. 
Microscopically, the bulk modulus measures the local curvature of the bond-dissociation curves in the bulk system.
Here, we benchmark the bulk moduli of 24 bulk samples using our built NAOs. 
The bulk modulus $B_0$ is calculated by fitting the Birch-Murnaghan equation~\cite{birchmurnaghan1947} (in Eqn.~\ref{eq:eos}), where $E$ is the energy, $V$ is the volume, $V_0$ is the equilibrium volume, and $B_0^\prime$ is the derivative of bulk modulus $B_0$ with respect to the pressure. 
In our tests, $V$ ranges from 96\% to 104\% with a step size of 2\% isotropically of $V_0$ obtained by a full relaxation with PW basis. 
\begin{widetext}
\begin{align}\label{eq:eos}
    E\left( V \right) =E_0+\frac{9V_0B_0}{16}\left\{ \left[ \left( \frac{V_0}{V} \right) ^{2/3}-1 \right] ^3B_{0}^{\prime}+
    \left[ \left( \frac{V_0}{V} \right) ^{2/3}-1 \right] ^2\left[ 6-\left( \frac{V_0}{V} \right) ^{2/3} \right] ^2 \right\} 
\end{align}
\end{widetext}

\begin{table*}[]
\centering
\caption{Bulk modulus of 26 systems calculated with NAOs, PW basis sets, and experimental data (in GPa).}
\label{tab:crystal-bulk-modulus}
\begin{tabular}{@{}lllllll@{}}
\toprule
Bulk & 
\multicolumn{1}{c}{pVDZ} & 
\multicolumn{1}{c}{pVTZ$^-$} & 
\multicolumn{1}{c}{pVTZ} & 
\multicolumn{1}{c}{pVQZ$^=$} & 
\multicolumn{1}{c}{PW} 
& \multicolumn{1}{c}{Expt.}
\\ \midrule
$\mathrm{AlAs}$ & 67  & 66   & 67   & 66   & 67  & 77\textsuperscript{a} \\
$\mathrm{AlP}$  & 82  & 83   & 83   & 82   & 82  & 86\textsuperscript{a} \\
$\mathrm{AlSb}$ & 49  & 49   & 49   & 50   & 49  & 58\textsuperscript{a} \\
$\mathrm{AlN}$  & 190 & 193  & 193  & 192  & 195 & 202\textsuperscript{b}, 208\textsuperscript{c}, 237\textsuperscript{d}, 160\textsuperscript{e}\\
$\mathrm{BN}$   & 369 & 370  & 369  & 369  & 370 & 465\textsuperscript{f}\\
$\mathrm{BP}$   & 160 & 161  & 161  & 161  & 160 & 173\textsuperscript{f}\\
C               & 434 & 432  & 434  & 431  & 432 & 442\textsuperscript{a} \\
$\mathrm{CdTe}$ & 36  & 35   & 35   & 35   & 35  & 42\textsuperscript{a} \\
$\mathrm{CdS}$  & 55  & 54   & 54   & 54   & 53  & 62\textsuperscript{a} \\
$\mathrm{CdSe}$ & 46  & 45   & 45   & 45   & 45  & 53\textsuperscript{a} \\
$\mathrm{GaAs}$ & 60  & 61   & 61   & 61   & 61  & 75\textsuperscript{a}, 75.57\textsuperscript{g}\\
$\mathrm{GaP}$  & 76  & 76   & 77   & 77   & 76  & 89\textsuperscript{a}\\
$\mathrm{GaSb}$ & 45  & 45   & 45   & 45   & 45  & 57\textsuperscript{a}\\
$\mathrm{GaN}$  & 172 & 172  & 172  & 171  & 172 & 188\textsuperscript{h}\\
$\mathrm{InP}$  & 59  & 59   & 59   & 59   & 59  & 71\textsuperscript{a}\\
$\mathrm{LiF}$  & 71  & 68   & 66   & 68   & 68  & 73--74.4\textsuperscript{i}, 66.2--68.5\textsuperscript{j}\\
$\mathrm{MgO}$  & 150 & 154  & 152  & 155  & 152 & 159.7\textsuperscript{k}, 160.5\textsuperscript{l}, 162.5$\pm$0.7\textsuperscript{m}\\
$\mathrm{MgS}$  & 76  & 75   & 75   & 75   & 75  & 79.8$\pm$0.37\textsuperscript{l}, 76.0$\pm$0.13\textsuperscript{l}, 81.4$\pm$0.29\textsuperscript{l}\\
$\mathrm{NaCl}$ & 24  & 24   & 24   & 24   & 24  & 23.9\textsuperscript{l}, 25.03$\pm$0.08\textsuperscript{n}\\
Si              & 88  & 87   & 88   & 88   & 87  & 98\textsuperscript{a}\\
$\mathrm{SiC}$  & 210 & 210  & 209  & 209  & 211 & 205\textsuperscript{o}, 237$\pm$2\textsuperscript{o}\\
$\mathrm{ZnSe}$ & 57  & 56   & 56   & 56   & 57  & 62\textsuperscript{a}\\
$\mathrm{ZnTe}$ & 44  & 43   & 43   & 43   & 43  & 51\textsuperscript{a}\\
$\mathrm{ZnO}$  & 129 & 129  & 129  & 131  & 130 & \\
$\mathrm{ZnO}$w & 127 & 129  & 129  & 131  & 130 & 139$\pm$8\textsuperscript{p}\\
$\mathrm{ZnS}$  & 71  & 70   & 70   & 69   & 70  & 77\textsuperscript{a}\\
                &     &      &      &      &     & \\
MAE$_\mathrm{PW}$ & 1.00   & 0.54    & 0.53    & 0.74    & & \\
MRE$_\mathrm{PW}$ & 1.13\% & 0.57\%  & 0.52\%  & 0.64\%  & & \\ 
\bottomrule
\end{tabular}

\medskip
\raggedright
\footnotesize
Experimental data from: \textsuperscript{a}Cohen~\cite{cohen1985bulkdataset}; \textsuperscript{b}Tsubouchi \textit{et al}.~\cite{tsubouchi1981AlNbulkmoduli1}; \textsuperscript{c}Ueno \textit{et al}.~\cite{ueno1992AlNbulkmoduli2}; \textsuperscript{d}McNeil \textit{et al}.~\cite{mcneil1993AlNbulkmoduli3}; \textsuperscript{e}Gerlich \textit{et al}.~\cite{gerlich1986AlNbulkmoduli4}; \textsuperscript{f}Wentzcovitch \textit{et al}.~\cite{wentzcovitch1986BNBPdataset}; \textsuperscript{g}Juan and Kaxiras~\cite{juan1993application}; \textsuperscript{h}Xia \textit{et al}.~\cite{xia1993GaNbulkmoduli}; \textsuperscript{i}Liu \textit{et al}.~\cite{liu2007LiFbulkmoduli1}; \textsuperscript{j}Yagi~\cite{yagi1978LiFbulkmoduli2}; \textsuperscript{k}Karki \textit{et al}.~\cite{karki1997MgObulkmoduli1}; \textsuperscript{l}Peiris \textit{et al}.~\cite{peiris1994MgSbulkmoduli}; \textsuperscript{m}Zha \textit{et al}.~\cite{zha2000MgObulkmoduli2}; \textsuperscript{n}Decker~\cite{decker1971NaClbulkmoduli}; \textsuperscript{o}Wang \textit{et al}.~\cite{wang2016SiCbulkmoduli}; \textsuperscript{p}Hanna \textit{et al}.~\cite{hanna2011ZnObulkmoduli}.
\end{table*}

As shown in Figure~\ref{fig:crystal-summary}(c), the errors of all NAOs distribute nearly symmetrically around 0, indicating no systematic error. 
Increasing the number of basis functions can further narrow the error distribution.
It can be observed from Table~\ref{tab:crystal-bulk-modulus} that the MAE and MRE substantially decrease with the increase of the number of basis functions, in a similar manner as other properties. 
Among NAOs, the highest MAE is 1 GPa; this value is near the precision of bulk modulus measurement, which indicates that all NAOs can have excellent precision.
pVQZ$^=$ is found to have a smaller average error while larger MRE and a significantly wider error distribution than pVTZ$^-$ (Figure~\ref{fig:crystal-summary}(c) and (h)).
A close examination reveals that the pVQZ$^=$ basis set mainly fails in cases AlN, BN, GaN, MgO, and SiC, which is also reported in Table IX. of Lin \textit{et al}.~\cite{Lin2021}, but has the smallest error in the cases of CdTe, CdS, GaAs, and ZnOw.
For the cases AlN, GaN, and MgO, we stress that, during compression and stretching of chemical bonds, the requirement for higher-angular-momentum basis functions to describe the wavefunction becomes dominant, which can also rationalize the improvement in precision from pVTZ$^-$ to pVTZ. 
For the cases BN and SiC, the relative error is relatively small; this behavior may primarily stem from the numerical conditioning of the spillage optimization problem.
For the CdTe, CdS, GaAs, and $\mathrm{ZnO}$w, on the contrary, because pVTZ would have g-type basis functions for Cd, Te, Ga, and Zn elements, which are less dominant in this test, an addition of the s/p/d/f-type basis functions therefore gains more profit in the accuracy.

We also introduce the $\Delta$-metric to quantify the difference between two EOS curves in the volume range from 96\% $V_0$ ($V_\mathrm{m}$) to 104\% $V_0$ ($V_\mathrm{M}$). 
With marks $A$ and $B$ distinguishing two EOS curves, the $\Delta \left( A,B \right)$ is defined as 
\begin{align}
    \Delta \left( A,B \right) =\frac{1}{V_{\mathrm{M}}-V_{\mathrm{m}}}\sqrt{\int_{V_{\mathrm{m}}}^{V_{\mathrm{M}}}{\left[ E^A\left( V \right) -E^B\left( V \right) \right] ^2\mathrm{d}V}}.~~ \label{eq:delta}
\end{align}
In Figure~\ref{fig:crystal-summary}(d) and Table~\ref{tab:crystal-delta}, the distribution and values of $\Delta$ between NAOs and PW are shown.

Among the $\Delta$ values of pVDZ, there are 35\% (9 cases) below 1 meV/atoms, 62\% (16 cases) below 1.5 meV/atoms. 
Such values are comparable with the 1 meV/atom threshold suggested by Prandini \textit{et al}.~\cite{prandini2018sssp} in the benchmark of pseudopotentials against the all-electron, where pseudopotentials with $\Delta <$ 1 meV/atom are considered to be reliable.
%\footnote{As reported, there are also cases where there is no pseudopotential that can have $\Delta <$ 1 meV/atom; in such cases, the one with the lowest is the recommended}.
When the NAO increases to pVTZ$^-$, 58\% of cases have $\Delta$ values below the 1 meV/atom. For pVTZ, there are only the LiF, ZnO, ZnOw, and ZnS that have $\Delta$ values still larger than 1 meV/atom, while for ZnO, its value is comparable with other atom-centered orbital-based DFT codes~\cite{bosoni2024acwf}.

By comparing the values of pVTZ$^-$, pVTZ, and pVQZ$^=$ in Table~\ref{tab:crystal-delta}, it can be found that the dominant origin of the error reduction varies from system to system. 
In the systems LiF, ZnO, and ZnOw, pVQZ$^=$ out-performs pVTZ, indicating that introducing additional basis functions with restricted angular momenta is more effective, whereas in the other systems, higher-angular-momentum basis functions contribute dominantly to the precision improvement.

\paragraph{Cohesive energies}
The cohesive energy, defined analogously to the atomization energy, is a critical indicator of basis set performance. 
In bulk systems, it also provides indirect insights into the relative stability of phases with identical composition—such as the zincblende and wurtzite polymorphs of ZnO — an attribute essential for constructing accurate phase diagrams and discovering novel materials. 
Here, we benchmark the cohesive energy of our NAOs against PW results across 26 systems to validate their ability to describe this key thermodynamic property.

\begin{table}[]
\centering
\caption{Cohesive energies of 26 bulk systems calculated with NAOs and PW basis sets (in eV/atom).}
\label{tab:crystal-cohesive-energies}
\begin{tabular}{@{}llllll@{}}
\toprule
Bulk & 
\multicolumn{1}{c}{pVDZ} & 
\multicolumn{1}{c}{pVTZ$^-$} & 
\multicolumn{1}{c}{pVTZ} & 
\multicolumn{1}{c}{pVQZ$^=$} & 
\multicolumn{1}{c}{PW} \\ \midrule
$\mathrm{AlAs}$ & 3.679 & 3.685 & 3.697 & 3.691 & 3.704 \\
$\mathrm{AlP}$  & 4.103 & 4.112 & 4.125 & 4.119 & 4.136 \\
$\mathrm{AlSb}$ & 3.349 & 3.353 & 3.360 & 3.359 & 3.364 \\
$\mathrm{AlN}$  & 5.591 & 5.615 & 5.644 & 5.633 & 5.669 \\
$\mathrm{BN}$   & 6.875 & 6.877 & 6.892 & 6.874 & 6.888 \\
$\mathrm{BP}$   & 5.336 & 5.346 & 5.366 & 5.344 & 5.362 \\
C               & 7.751 & 7.740 & 7.753 & 7.746 & 7.727 \\
$\mathrm{CdTe}$ & 2.191 & 2.085 & 2.088 & 2.089 & 2.089 \\
$\mathrm{CdS}$  & 2.748 & 2.625 & 2.634 & 2.618 & 2.636 \\
$\mathrm{CdSe}$ & 2.466 & 2.342 & 2.353 & 2.340 & 2.358 \\
$\mathrm{GaAs}$ & 3.153 & 3.147 & 3.155 & 3.149 & 3.154 \\
$\mathrm{GaP}$  & 3.509 & 3.505 & 3.514 & 3.508 & 3.515 \\
$\mathrm{GaSb}$ & 2.950 & 2.939 & 2.942 & 2.941 & 2.939 \\
$\mathrm{GaN}$  & 4.300 & 4.300 & 4.309 & 4.303 & 4.309 \\
$\mathrm{InP}$  & 3.138 & 3.147 & 3.155 & 3.150 & 3.157 \\
$\mathrm{LiF}$  & 4.399 & 4.431 & 4.437 & 4.462 & 4.470 \\
$\mathrm{MgO}$  & 5.290 & 5.121 & 5.139 & 5.128 & 5.141 \\
$\mathrm{MgS}$  & 3.960 & 3.784 & 3.804 & 3.782 & 3.802 \\
$\mathrm{NaCl}$ & 3.145 & 3.143 & 3.146 & 3.149 & 3.153 \\
Si              & 4.579 & 4.585 & 4.602 & 4.590 & 4.601 \\
$\mathrm{SiC}$  & 6.385 & 6.394 & 6.436 & 6.411 & 6.432 \\
$\mathrm{ZnSe}$ & 2.757 & 2.603 & 2.614 & 2.605 & 2.629 \\
$\mathrm{ZnTe}$ & 2.409 & 2.271 & 2.273 & 2.276 & 2.280 \\
$\mathrm{ZnO}$  & 3.810 & 3.660 & 3.664 & 3.676 & 3.692 \\
$\mathrm{ZnO}$w & 3.802 & 3.653 & 3.656 & 3.668 & 3.684 \\
$\mathrm{ZnS}$  & 3.114 & 2.960 & 2.969 & 2.959 & 2.985 \\
                &       &       &       &       &        \\
MAE$_\mathrm{PW}$       & 0.064 & 0.018 & 0.009 & 0.013 &        \\
MRE$_\mathrm{PW}$       & 1.95\% & 0.47\% & 0.24\% & 0.34\% &        \\ \bottomrule
\end{tabular}
\end{table}

In Figure~\ref{fig:crystal-summary}(e), the distribution of error of pVDZ spans about 0.23 eV/atom wide. 
By increasing the number of basis functions to pVTZ$^-$, the distribution narrows rapidly to about 0.06 eV/atom.
This behavior indicates that the highly precise cohesive energy calculation would pose a strong demand on the basis transferability to describe distinct chemical environments (isolated atoms and close-packed bulks). Therefore, in alignment with the atomization energy benchmark, the pVTZ contains basis functions with higher angular momentum, which can perform better than the pVQZ$^=$.  
From Table~\ref{tab:crystal-cohesive-energies}, it can be observed that NAOs larger than pVDZ can have MAE smaller than 0.042 eV/atom, which satisfies the chemical accuracy, and in Figure~\ref{fig:crystal-summary}(e), most values of pVTZ lie between the (-0.042, 0.042) eV/atom.

\paragraph{Bandgap}
The bandgap (or energy gap) is defined as the energy difference between the valence band maximum (VBM) and conduction band minimum (CBM). 
It serves as a critical benchmark for validating the accuracy of electronic-structure calculations, as it can be a direct indicator of the material's electronic properties and plays a crucial role in electronic-transport calculations. 
Accurate prediction of bandgaps is essential for guiding material discovery, particularly in applications such as semiconductors and photovoltaics. 
Here, we benchmark our NAOs against PW in terms of bandgap prediction precision.

\begin{table}[]
\centering
\caption{Bandgap benchmark on 26 bulk systems (in eV).}
\label{tab:crystal-bandgap}
\begin{tabular}{@{}llllll@{}}
\toprule
Bulk & 
\multicolumn{1}{c}{pVDZ} & 
\multicolumn{1}{c}{pVTZ$^-$} & 
\multicolumn{1}{c}{pVTZ} & 
\multicolumn{1}{c}{pVQZ$^=$} & 
\multicolumn{1}{c}{PW} \\ \midrule
$\mathrm{AlAs}$ & 1.48 & 1.47 & 1.47 & 1.48 & 1.47 \\
$\mathrm{AlP}$  & 1.65 & 1.65 & 1.65 & 1.64 & 1.64 \\
$\mathrm{AlSb}$ & 1.25 & 1.24 & 1.24 & 1.24 & 1.24 \\
$\mathrm{AlN}$  & 4.09 & 4.11 & 4.07 & 4.10 & 4.08 \\
$\mathrm{BN}$   & 4.56 & 4.54 & 4.54 & 4.54 & 4.54 \\
$\mathrm{BP}$   & 1.30 & 1.29 & 1.28 & 1.28 & 1.28 \\
C               & 4.24 & 4.18 & 4.18 & 4.18 & 4.17 \\
$\mathrm{CdTe}$ & 0.56 & 0.57 & 0.57 & 0.57 & 0.58 \\
$\mathrm{CdS}$  & 1.10 & 1.11 & 1.11 & 1.11 & 1.11 \\
$\mathrm{CdSe}$ & 0.52 & 0.53 & 0.52 & 0.53 & 0.52 \\
$\mathrm{GaAs}$ & 0.52 & 0.54 & 0.54 & 0.54 & 0.54 \\
$\mathrm{GaP}$  & 1.62 & 1.59 & 1.59 & 1.57 & 1.56 \\
$\mathrm{GaSb}$ & 0.34 & 0.33 & 0.33 & 0.33 & 0.32 \\
$\mathrm{GaN}$  & 1.96 & 1.76 & 1.75 & 1.75 & 1.76 \\
$\mathrm{InP}$  & 0.46 & 0.46 & 0.45 & 0.43 & 0.43 \\
$\mathrm{LiF}$  & 8.63 & 8.74 & 8.76 & 8.86 & 8.90 \\
$\mathrm{MgO}$  & 4.44 & 4.44 & 4.46 & 4.44 & 4.47 \\
$\mathrm{MgS}$  & 2.73 & 2.73 & 2.77 & 2.75 & 2.78 \\
$\mathrm{NaCl}$ & 4.96 & 4.96 & 4.97 & 4.97 & 4.99 \\
Si              & 0.62 & 0.63 & 0.62 & 0.62 & 0.61 \\
$\mathrm{SiC}$  & 1.46 & 1.39 & 1.40 & 1.39 & 1.40 \\
$\mathrm{ZnSe}$ & 1.11 & 1.13 & 1.13 & 1.13 & 1.13 \\
$\mathrm{ZnTe}$ & 1.06 & 1.07 & 1.07 & 1.06 & 1.07 \\
$\mathrm{ZnO}$  & 0.82 & 0.81 & 0.81 & 0.80 & 0.81 \\
$\mathrm{ZnO}$w & 0.72 & 0.70 & 0.70 & 0.69 & 0.70 \\
$\mathrm{ZnS}$  & 2.01 & 2.02 & 2.02 & 2.01 & 2.01 \\
        &       &         &         &          &        \\
MAE$_\mathrm{PW}$ & 0.04    & 0.02    & 0.01    & 0.01   &        \\
MRE$_\mathrm{PW}$ & 2.40\%  & 1.03\%  & 0.69\%  & 0.59\%  &        \\ \bottomrule
\end{tabular}
\end{table}

As shown in Figure~\ref{fig:crystal-summary}(f), all NAOs have average bandgap error close to 0, and in Table~\ref{tab:crystal-bandgap}, all NAOs have MAE not larger than 0.04 eV, which are near or below the standard deviation of the experimental bandgap measurement~\cite{bandgapmeasure2018,taucmethod2015}.
From Table~\ref{tab:crystal-bandgap} and Figure~\ref{fig:crystal-summary}(f), it can be seen that the introduction of additional basis functions with restricted angular momenta firstly eliminates the error of the GaN case, suppresses the positive tail of the pVDZ violin, and leaves the LiF still at the end of the negative tail.
The similar lengths of the negative tails of pVTZ$^-$ and pVTZ indicates the higher angular momentum basis functions do not contribute significantly in improving the bandgap prediction, this is reasonable because the cases where the atomic orbitals with angular momenta $l+2$ (where $l$ is the largest angular momentum of electron in atomic configuration) the contributes significantly to states near the VBM or CBM are always scarcely seen.
By leveling up the pVTZ$^-$ to pVQZ$^=$, the error of LiF is reduced to 0.04 eV, and the negative tail of the violin is largely suppressed.

\paragraph{$\eta$-test}
To extensively benchmark the precision of the electronic structure in bulk systems, the $\eta$-metrics is employed in this section, where the summation over eigenstates (indexed by $n$ in eqn.~\ref{eqn:eta}) is extended to include $\mathbf{k}$-points. 
A precise band-structure calculation is a cornerstone for understanding numerous properties, including transport and optical properties of solids.
Here, we perform a band structure alignment benchmark using the $\eta$-metrics (eqn.~\ref{eqn:eta}) on NAOs.

It is shown in Figure~\ref{fig:crystal-summary}(g) and Table~\ref{tab:crystal-eta} that the $\eta$ value decreases with the increase of radial functions, from pVDZ to pVTZ or pVQZ$^=$. The inclusion of high angular momentum basis functions is found to be capable of improving the precision of the occupied band structure calculation to some degree; as a result, the distribution of $\eta$ in the range (0.03, 0.06) has been suppressed.
On the other hand, the pVTZ is still deficient compared with the pVQZ$^=$ basis sets in LiF, most significantly
%, and slightly in CdTe and ZnSe, in which there are semi-core states (Li, Cd, and Zn) whose descriptions are better with pVQZ$^=$
, coinciding with the bandgap benchmarks. 

To evaluate NAO precision in a wider energy range, including more conduction bands, another metric $\eta_{10}$ that is derived from $\eta$ by manually lifting the Fermi level by 10 eV in the $\eta$ metric evaluation (which in turn causes an increase in the number of electrons of the system) is introduced. 
We also generate two new NAOs with notations distinguishing from those whose spillage only includes the occupied states were introduced here, pVTZ-2V and pVTZ-3V, which represent including virtual states by twice or three times of the number of occupied states into the spillage during the construction of pVTZ basis sets based on the pVDZ, respectively, that is meaningful in ensuring the practical validity of strategy of improving the transferability of basis by means of including more reference states.

\begin{figure}
    \centering
    \includegraphics[width=1.0\linewidth]{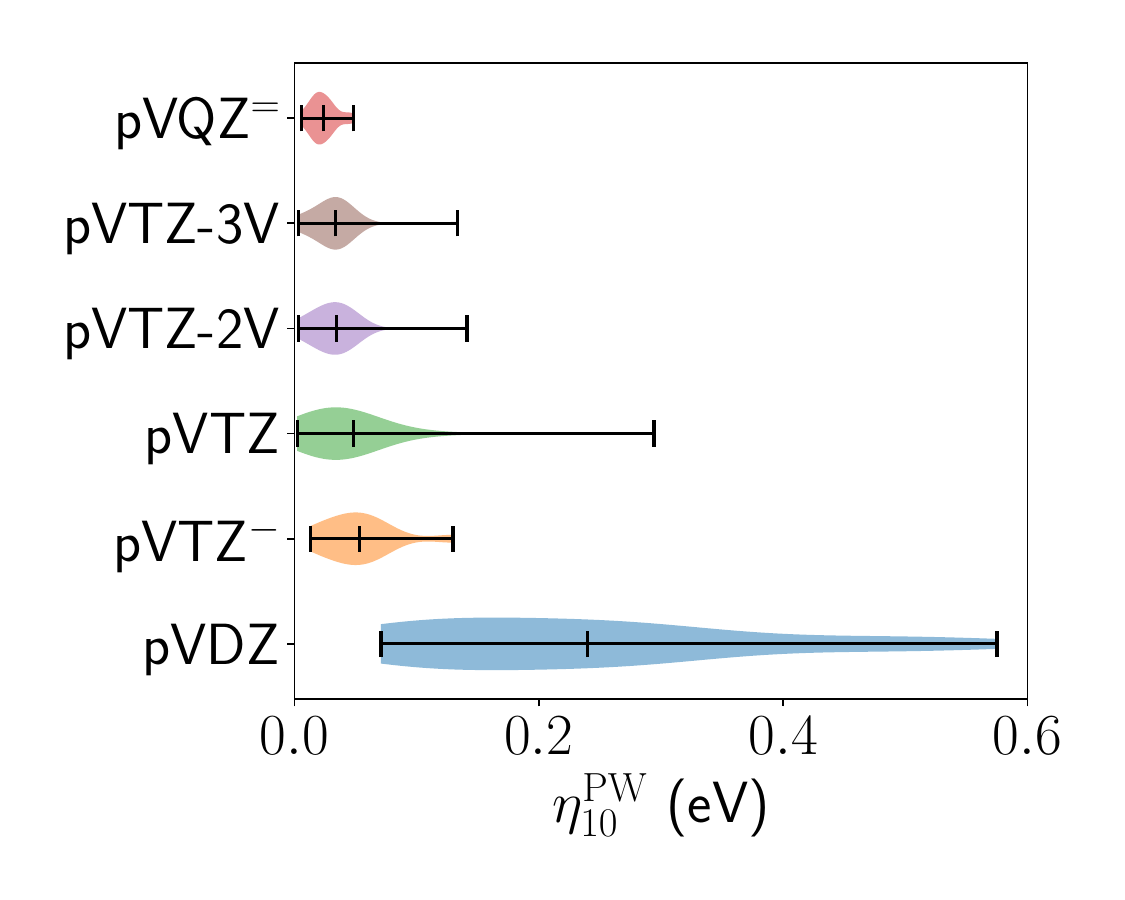}
    \caption{Distribution of $\eta_{10}$ metrics with respect to PW (denoted as $\eta^\mathrm{PW}_{10}$) of pVDZ (blue), pVTZ$^-$ (orange), pVTZ (green), pVTZ-2V (purple), pVTZ-3V (brown) and pVQZ$^=$ (red) basis sets, from the bottom to the top, respectively. The extrema and average are indicated with black whiskers. Detailed data can be found in Table~\ref{tab:crystal-eta10}.}
    \label{fig:crystal-eta10-violinplot}
\end{figure}

Figure~\ref{fig:crystal-eta10-violinplot} shows that the distribution of $\eta^\mathrm{PW}_{10}$ is wider than $\eta^\mathrm{PW}$, for example the $\eta^\mathrm{PW}$ distribution of pVDZ spans with width about 0.1 eV, while $\eta^\mathrm{PW}_{10}$ has the width about 0.5 eV. 
Such behavior indicates that additional construction or modification of NAOs is necessary to describe high-lying conduction bands accurately. 
By comparing the pVTZ$^-$ with pVTZ, it is found that by including the f-type radial function into basis sets, there are cases whose $\eta^\mathrm{PW}_{10}$ values decrease to nearly zero, while at cost, the distribution of $\eta^\mathrm{PW}_{10}$ becomes much wider, which implies the f-type radial function generated by only including occupied reference states can not have enough transferability to describe the states in conduction bands.
Comparatively, by including virtual states (pVTZ vs. pVTZ-2V/pVTZ-3V), the distribution of $\eta^\mathrm{PW}_{10}$ narrows significantly, the average of $\eta^\mathrm{PW}_{10}$ also decreases. 
These results unambiguously demonstrate that improving transferability can be achieved by including more states.
However, in Figure~\ref{fig:crystal-eta10-violinplot}, we note the pVQZ$^=$ basis set still has the narrowest $\eta^\mathrm{PW}_{10}$ distribution and smallest averaged $\eta^\mathrm{PW}_{10}$.
In Table~\ref{tab:crystal-eta10} it can be found that the pVQZ$^=$ basis can have a lower $\eta_{10}$ value in most systems except AlN, Si (diamond), C (diamond), SiC, BP, BN and AlAs, in which the pVTZ-2V or pVTZ-3V has smaller $\eta^\mathrm{PW}_{10}$.
Among all bulk systems tested, NaCl has the largest $\eta_{10}$ of pVTZ, pVTZ-2V, and pVTZ-3V. 
For the GaN system, the inclusion of virtual states into the spillage even degrades precision (the $\eta_{10}$ value increases from 0.0069 to 0.0160 eV for pVTZ-2V and to 0.0246 for pVTZ-3V). 
To obtain a more direct understanding of the effect of including virtual states in spillage and how this improves the description of conduction bands, and to identify the origin of the error, band-structure calculations are performed for these two systems. 
The $\mathbf{k}$-point paths are generated with the SeeK-path toolkit~\cite{hinuma2017band, togo2024spglib}.

\begin{figure*}
    \centering
    \includegraphics[width=1.0\linewidth]{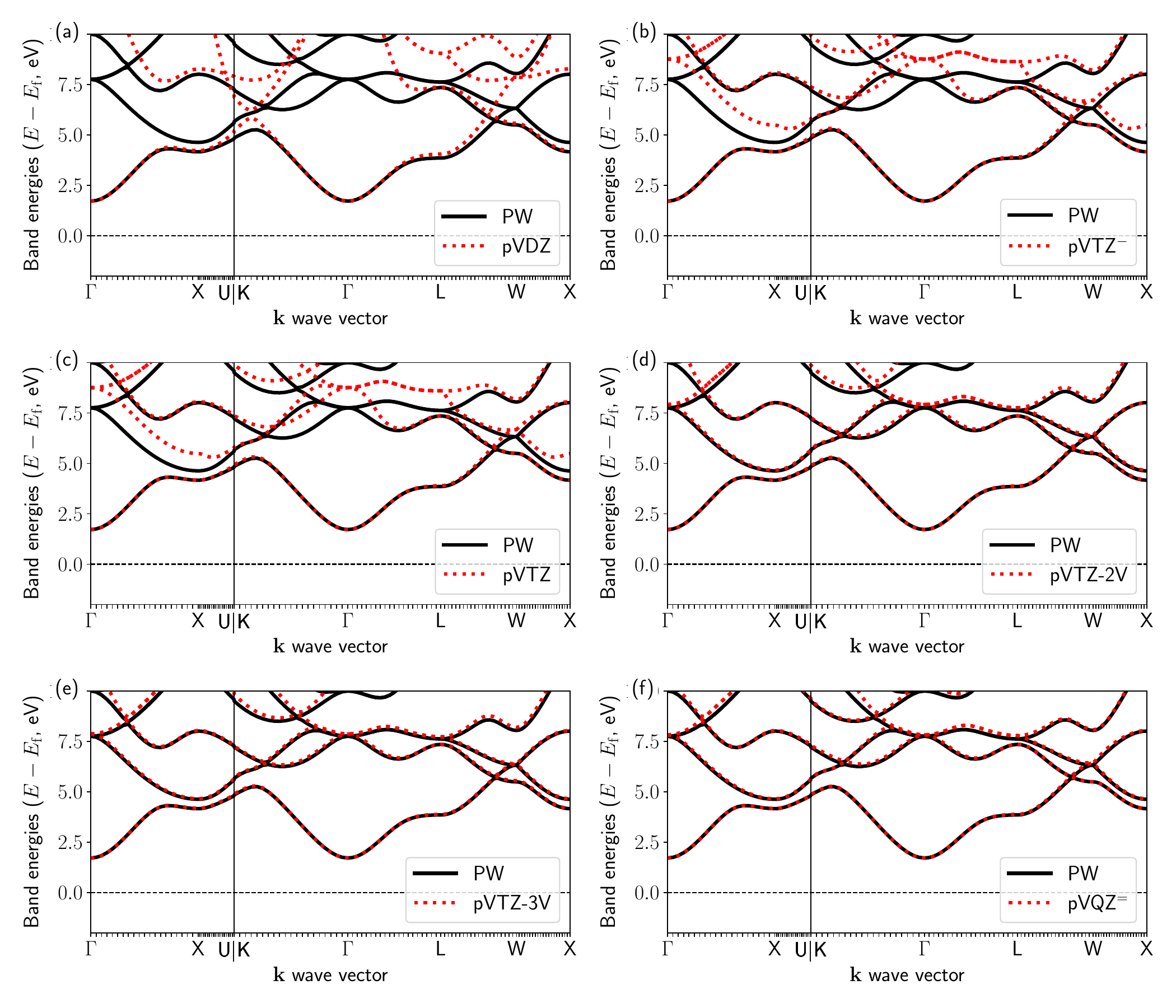}
    \caption{Comparison on the conduction band structure of NaCl between PW and NAOs, including (a) pVDZ, (b) pVTZ$^-$, (c) pVTZ, (d) pVTZ-2V, (e) pVTZ-3V, and (f) pVQZ$^=$. All band structures calculated with PW are plotted with black solid lines, and those of NAOs are plotted with red dotted lines. For NAOs in (a-c) and (f), as mentioned in the text, only occupied states are included in spillage during generation; for NAOs in (d, e), occupied states and additional virtual states are included.}
    \label{fig:crystal-nacl-bandstructure}
\end{figure*}

Figure~\ref{fig:crystal-nacl-bandstructure} shows the band structures of NaCl calculated with various NAOs, where it is found that all NAOs can have the CBM that coincides with PW well, which shows agreement with the bandgap benchmark that the largest error of NaCl bandgap prediction is only 0.03 eV. 
NAOs larger than pVDZ can have qualitatively correct prediction at all the $\mathbf{k}$-points on the lowest conduction band, and pVTZ$^-$ and pVTZ fail on the majority of $\mathbf{k}$-points on the higher conduction bands.
In Figure~\ref{fig:crystal-nacl-bandstructure} (d) and (e), it can be seen that the strategy of including virtual states into spillage to improve the precision is still valid, it can result in qualitatively correct band structure prediction up to the energy above Fermi level for at least 10 eV, while at a large number of $\mathbf{k}$-points, the eigenvalues calculated with pVQZ$^=$ are indeed lower than pVTZ series NAOs and the band structure of pVQZ$^=$ coincide with PW results the most.
Together with the result that pVTZ$^-$ has nearly identical band structure with pVTZ, we stress that the large $\eta_{10}$ is because the conduction bands within the energy range up to about 10 eV above the Fermi level are contributed by s, p, and d orbitals.

\begin{figure*}
    \centering
    \includegraphics[width=1.0\linewidth]{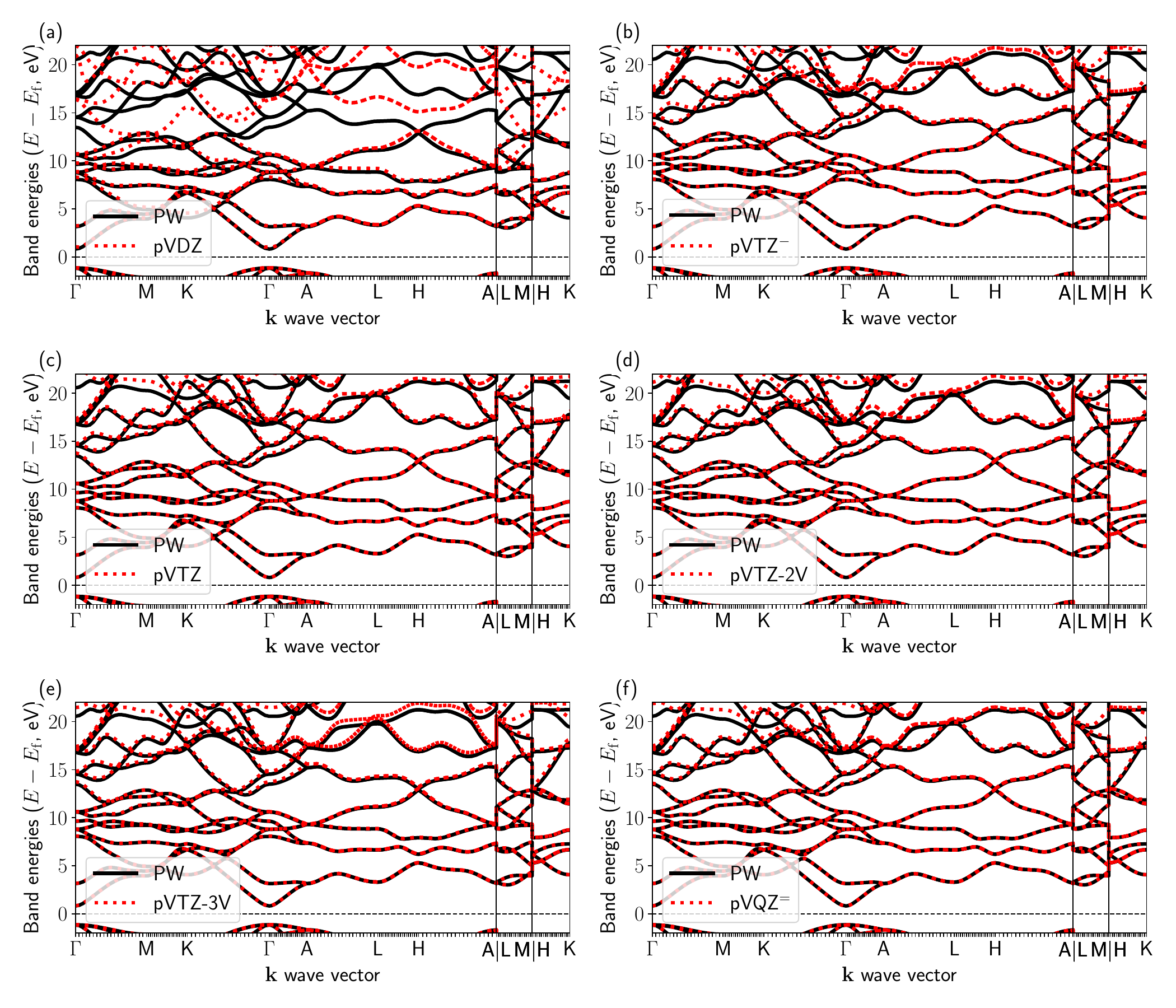}
    \caption{Comparison on the conduction band structure of GaN between PW and NAOs, including (a) pVDZ, (b) pVTZ$^-$, (c) pVTZ, (d) pVTZ-2V, (e) pVTZ-3V, and (f) pVQZ$^=$. All band structures calculated with PW are plotted with black solid lines, and those of NAOs are plotted with red dotted lines. For NAOs in (a-c) and (f), as mentioned in the text, only occupied states are included in spillage during generation; for NAOs in (d, e), occupied states and additional virtual states are included.}
    \label{fig:crystal-gan-bandstructure}
\end{figure*}

GaN is a III/V direct bandgap semiconductor commonly used in blue light-emitting diodes. 
We checked the band structure over a wider energy range, corresponding to excitation wavelengths below 50 nm, thereby covering most of the optical spectrum. 
As shown in Figure~\ref{fig:crystal-gan-bandstructure} (a), it is similar to the NaCl case that only pVDZ overestimates the eigenvalues at some $\mathbf{k}$-point on the lowest conduction band.
When the basis set increases to pVTZ$^-$, there is no qualitative difference with PW up to 10 eV above the Fermi level. 
Inclusion of f-type orbital for N and g-type for Ga slightly lowers the bands in the range from 15 to 20 eV above the Fermi level, bringing a 0.006 eV decrease in $\eta_{10}$.
The additional inclusion on the virtual states into spillage generates the pVTZ-2V and 3V, improves part of precision, as depicted in Figure~\ref{fig:crystal-gan-bandstructure} (d) and (e), for high-lying states at about 20 eV above the Fermi level, on the path of kpoints $\Gamma - \mathrm{M} - \mathrm{K} - \Gamma$, pVTZ-2V and sometimes the pVTZ-3V can have energy closer to PW than pVTZ and pVQZ$^=$, but the contrary result can also be observed on the path $\mathrm{A} - \mathrm{L} - \mathrm{H} - \mathrm{A}$, this explains the rise in $\eta_{10}$ from pVTZ to pVTZ-2V and pVTZ-3V.

Therefore, although the $\eta_{10}$ values of pVTZ-2V and pVTZ-3V are less satisfying than pVQZ$^=$ in NaCl and GaN cases, they can still have qualitatively correct predictions on band structures and are still reliable in practical calculations.

To summarize, in bulk systems, pVDZ can always achieve acceptable precision in structural property prediction, whereas pVTZ and pVQZ$^=$ can achieve accurate predictions for all properties we tested.
The $\eta_{10}$ metrics and band-structure calculations support the idea that including virtual states in the spillage can enhance the transferability of NAOs without increasing the number of basis functions.
On the other hand, as the number of basis functions increases, the spillage optimization problem quickly becomes ill-conditioned and requires sophisticated non-convex optimization techniques.
Enlarging the band range included makes the strategy of increasing the number of basis functions to improve the precision of the basis sets practically substantial.

\section{Summary}
In this work, we present a scheme for constructing NAO basis sets via contraction of TSWs. 
The contraction is obtained by minimizing the trace of the kinetic operator in the residual space, which is equivalent to a generalization of the spillage function-minimizing scheme.
Besides our NAOs possess systematical improvability that can approach to the CBS by increasing the truncation radius, extending the set of angular momenta included and adding more radial functions, we also demonstrate a strategy for determining the truncation radius and the maximal angular momentum included of the NAOs, by converging the energy error of homonuclear di-atomic molecule against the near-complete PW with given threshold, by which the completeness and accuracy of NAOs are significantly improved. 
Our method can eliminate spurious interactions between periodic images and distorted information that may degrade the NAOs. 
Further enables the transferability-improvement strategy by including virtual states in the spillage, which is effective for conduction-band calculations.
Benchmarks show that, in both molecular and bulk systems, the constructed NAOs exhibit good precision for properties such as total energy, bond length, atomization energy, lattice constant, cohesive energy, and band gap.

\section{Code availability}
ABACUS is open-sourced on the GitHub repository (https://github.com/deepmodeling/abacus-develop).

The NAO generation implementation is open-sourced on GitHub under the GPL 3.0 license (https://github.com/MCresearch/ABACUS-CSW-NAO).

High-throughput benchmark is performed with the help of the open-sourced APNS (ABACUS-Pseudopotential-Numerical atomic orbital-Square) workflow (https://github.com/kirk0830/ABACUS-Pseudopot-Nao-Square).

\section{Acknowledgement}
This work is supported by the ABACUS Pseudopotential Numerical Atomic Orbital project of AI for Science Institute. M.C is supported by National Key R$\&$D Program of China under Grant No. 2025YFB3003603. We thank Lixin He from the University of Science and Technology of China (USTC), Peize Lin from Institute of Artificial Intelligence, Hefei Comprehensive National Science Center, and Xinguo Ren from the Institute of Physics (IOP), Chinese Academy of Sciences (CAS) for the suggestions on NAOs benchmark and discussion on results, and Weiqing Zhou from Wuhan University, China for the technical advice on the SCF convergence.

\section{Appendix}

\appendix

\section{Additional notes on improving the NAOs transferability by means of including virtual states into spillage}
\label{appendix:spurious-interaction-across-vacuum}
\renewcommand{\thetable}{A.\arabic{table}}
\renewcommand{\thefigure}{A.\arabic{figure}}
\setcounter{table}{0}
\setcounter{figure}{0}
The inclusion of virtual states into spillage enhances basis-set transferability, as shown in Figure~\ref{fig:crystal-eta10-violinplot} and Table~\ref{tab:crystal-eta10}; however, in our prior studies, we found that this approach fails when reference states are solved with PW as the expansion basis.
\begin{figure*}
    \centering
    \includegraphics[width=1.0\linewidth]{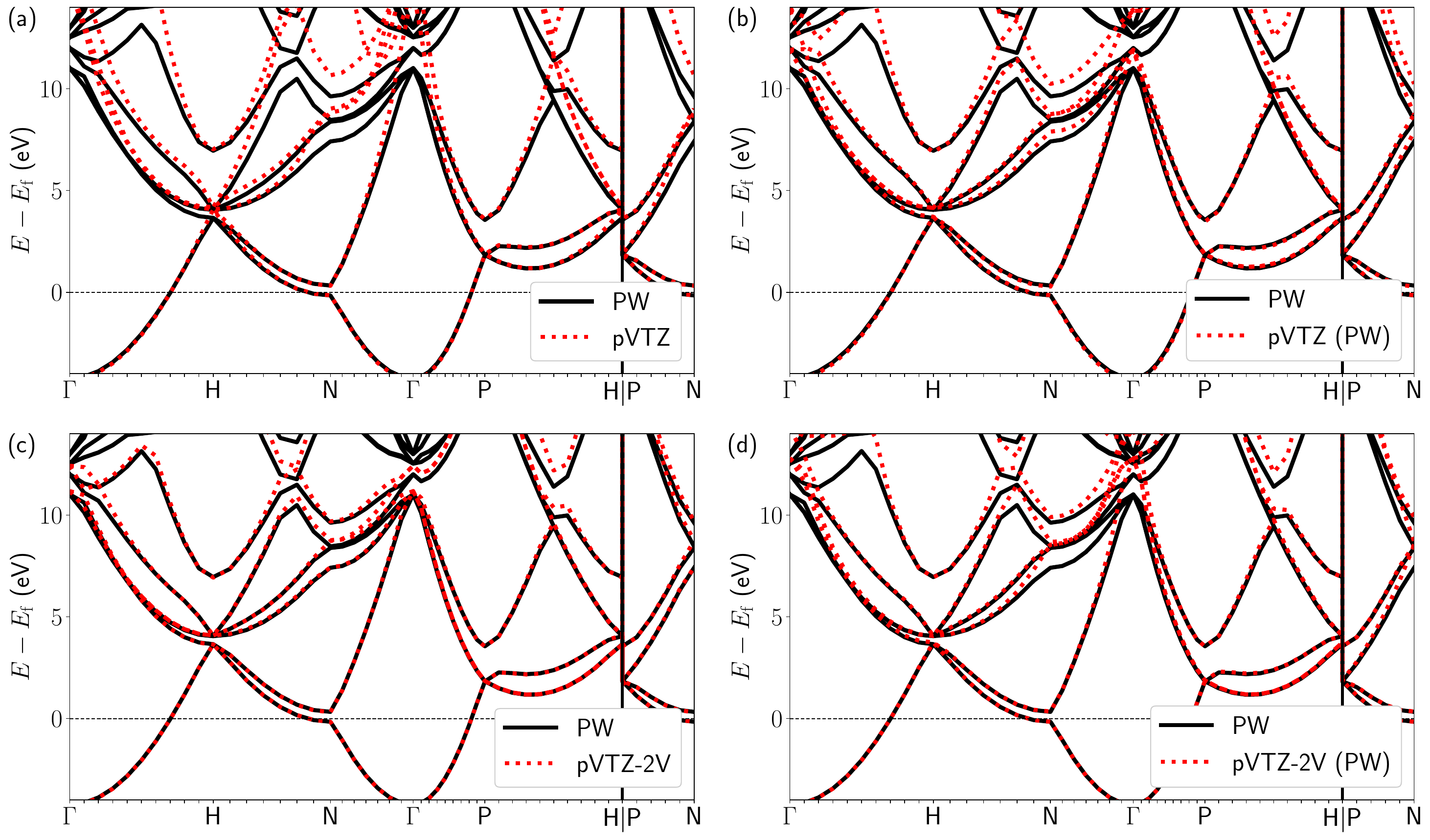}
    \caption{Na body-centered-cubic band structures calculated with PW, pVTZ, and pVTZ-2V basis sets. Specifically, the pVTZ and pVTZ-2V generated with PW-expanded reference states are marked with the additional suffix ``PW" to make them distinguishable from NAOs generated from the NSW contraction. In this test, the truncation radius $r_\mathrm{c}$ of NAOs is kept at 8 au. (a) Band structures calculated with PW (black solid line) and pVTZ basis (red dotted line). (b) band structures calculated with PW (black solid line) and pVTZ(PW) basis (red dotted line). (c) band structures calculated with PW (black solid line) and pVTZ-2V basis (red dotted line). band structures calculated with PW (black solid line) and pVTZ-2V(PW) basis (red dotted line). }
    \label{fig:sodium-jyvspw-bandstructure}
\end{figure*}

Figure~\ref{fig:sodium-jyvspw-bandstructure} compares the Na body-centered cubic crystal (BCC) band structures predicted with pVTZ and pVTZ-2V basis sets, whose reference states are expanded with NSW or PW (NAOs generated with PW-expanded reference states are marked with additional suffix ``PW'').
Taking PW (black solid line) as the reference, it shown that in the energy range around 10 eV higher than the Fermi level ($E_\mathrm{f}$), the pVTZ(PW) in Figure~\ref{fig:sodium-jyvspw-bandstructure}(b) has better description on conduction bands compared with pVTZ in Figure~\ref{fig:sodium-jyvspw-bandstructure}(a), especially near the $\Gamma$ point, and the path $\mathrm{P}-\mathrm{H}$. 
Conversely, pVTZ-2V in Figure~\ref{fig:sodium-jyvspw-bandstructure}(c) performs better than pVTZ-2V(PW) in Figure~\ref{fig:sodium-jyvspw-bandstructure}(d), the band prediction near the $\Gamma$, $\mathrm{N}$ points, and the path $\mathrm{P}-\mathrm{H}$ have been significantly improved.

\begin{figure}
    \centering
    \includegraphics[width=1.0\linewidth]{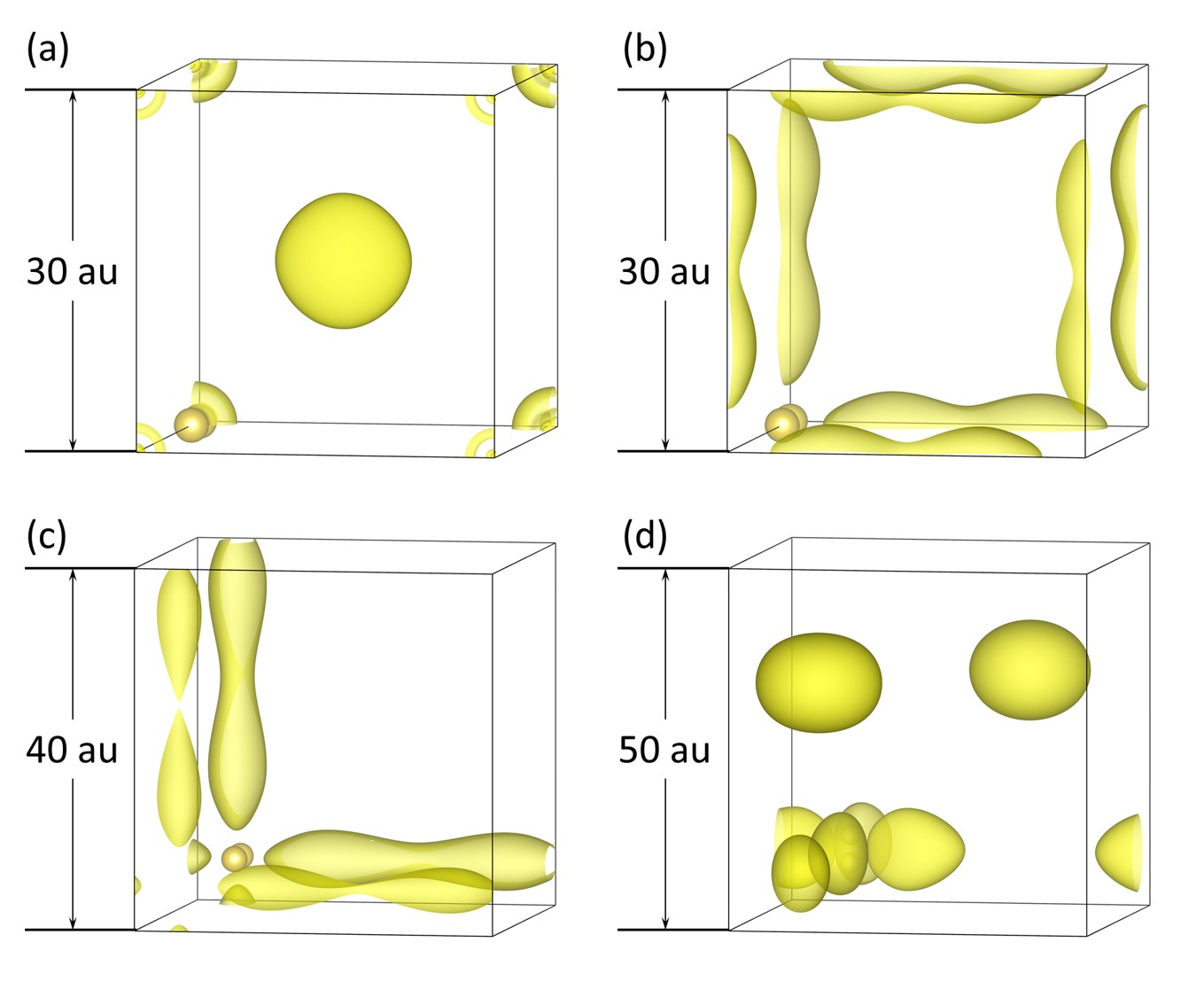}
    \caption{Non-physical virtual states (shown with $|\psi_{n\mathbf{k}}(\mathbf{r})|$) of $\mathrm{Na}_2$ molecule originated in different sizes of cubic simulation boxes from the unexpected overlap with periodic images. Molecule is always placed at the bottom right corner of the box, and atoms are colored with gold. (a, b) Two non-physical virtual states emerge in the box with edge length 30 au that are among the states included in spillage in the pVTZ-2V case. (c) A non-physical virtual state emerges in the box with edge length 40 au, which is within the same range as (a,b). (d) non-physical virtual state emerges in the box with edge length 50 au that is among the states included in the spillage in the pVTZ-3V case.}
    \label{fig:sodium-nonphysical-states}
\end{figure}

Such differences in transferability improvement arise from the inclusion of non-physical states (Figure~\ref{fig:sodium-nonphysical-states}) into spillage during the generation of pVTZ-2V(PW). 
These states originate from the overlap between periodic images of highly delocalized states. 
Similar phenomena can also be observed in the atomic PW calculation, where virtual states may exhibit an even number of degeneracies.
However, this kind of overlapping behavior of reference states in the PW case is not always consistent with the truncated primitive basis (TSW or NSW).
%, like what has been illustrated in Figure~\ref{fig:schematic-figure-csw-nao}(b), the overlap can be inaccessible in truncated basis cases because of the limitation of $r_\mathrm{c}$. 
As a result, those highly delocalized states are irreproducible in both shape and energy, making them less meaningful for improving the description of band structures in the range of interest. 
Although it is expected that these artifacts can be avoided by upscaling the simulation box, the high-scaling computational costs, uncertainties of state locality, and state ordering will drive this solution to be prohibitive in practice.

Compared with the truncation of NSW, the truncation eliminates the long tails of delocalized states and is similar to imposing a spherically symmetric confinement potential. 
As a result, all states solved with NSW are accessible to NAOs for construction. 
Although on the other side of the coin, the energies (eigenvalues) of these states are higher than the PW-expanded cases, the improvement in transferability is efficient, as shown in Figure~\ref{fig:sodium-jyvspw-bandstructure}.

\section{Additional benchmark data tables}
\renewcommand{\thetable}{B.\arabic{table}}
\setcounter{table}{0}
This section contains detailed NAOs benchmark data, including the $\eta$-metrics and $\Delta$ values.

The $\eta$ metric measures the energy level consistency between two methods; its definition has been introduced in eqn.~\ref{eqn:eta}. Here, the detailed values of $\eta$ between (PW, NSW) and NAOs are listed (Table~\ref{tab:molecule-eta}).
\begin{table*}[]
\centering
\caption{Occupation weighted energy levels differences ($\eta$-metrics) of NAOs with NSW and PW basis calculated on 11 molecule systems (in eV).}
\label{tab:molecule-eta}
\begin{tabular}{@{}lrrrrrrrrrrrrrrrrr@{}}
\toprule
\multirow{2}{*}{Molecule} &
  \multicolumn{3}{c}{pVDZ} &
  \multicolumn{1}{c}{\quad} &
  \multicolumn{3}{c}{pVTZ$^-$} &
  \multicolumn{1}{c}{\quad} &
  \multicolumn{3}{c}{pVTZ} &
  \multicolumn{1}{c}{\quad} &
  \multicolumn{3}{c}{pVQZ$^=$} &
  \multicolumn{1}{c}{\quad} &
  \multicolumn{1}{c}{NSW} \\ \cmidrule(lr){2-4} \cmidrule(lr){6-8} \cmidrule(lr){10-12} \cmidrule(lr){14-16} \cmidrule(l){18-18} 
 &
  \multicolumn{1}{c}{$\eta^\mathrm{NSW}$} &
  \multicolumn{1}{c}{\quad} &
  \multicolumn{1}{c}{$\eta^\mathrm{PW}$} &
  \multicolumn{1}{c}{} &
  \multicolumn{1}{c}{$\eta^\mathrm{NSW}$} &
  \multicolumn{1}{c}{\quad} &
  \multicolumn{1}{c}{$\eta^\mathrm{PW}$} &
  \multicolumn{1}{c}{} &
  \multicolumn{1}{c}{$\eta^\mathrm{NSW}$} &
  \multicolumn{1}{c}{\quad} &
  \multicolumn{1}{c}{$\eta^\mathrm{PW}$} &
  \multicolumn{1}{c}{} &
  \multicolumn{1}{c}{$\eta^\mathrm{NSW}$} &
  \multicolumn{1}{c}{\quad} &
  \multicolumn{1}{c}{$\eta^\mathrm{PW}$} &
  \multicolumn{1}{c}{} &
  \multicolumn{1}{c}{$\eta^\mathrm{PW}$} \\ \midrule
$\mathrm{Br}_2$ & 0.047 &  & 0.047 &  & 0.041 &  & 0.041 &  & 0.003 &  & 0.003 &  & 0.041 &  & 0.041 &  & 0.000 \\
$\mathrm{CO}$   & 0.046 &  & 0.048 &  & 0.028 &  & 0.030 &  & 0.012 &  & 0.014 &  & 0.022 &  & 0.023 &  & 0.003 \\
$\mathrm{Cl}_2$ & 0.050 &  & 0.050 &  & 0.043 &  & 0.043 &  & 0.004 &  & 0.005 &  & 0.043 &  & 0.043 &  & 0.000 \\
$\mathrm{F}_2$  & 0.025 &  & 0.025 &  & 0.024 &  & 0.025 &  & 0.004 &  & 0.004 &  & 0.024 &  & 0.024 &  & 0.000 \\
$\mathrm{I}_2$  & 0.079 &  & 0.079 &  & 0.026 &  & 0.026 &  & 0.009 &  & 0.009 &  & 0.025 &  & 0.025 &  & 0.000 \\
$\mathrm{LiH}$  & 0.061 &  & 0.061 &  & 0.040 &  & 0.040 &  & 0.041 &  & 0.041 &  & 0.012 &  & 0.012 &  & 0.000 \\
$\mathrm{Li}_2$ & 0.008 &  & 0.007 &  & 0.007 &  & 0.006 &  & 0.007 &  & 0.006 &  & 0.003 &  & 0.002 &  & 0.000 \\
$\mathrm{N}_2$  & 0.038 &  & 0.039 &  & 0.026 &  & 0.027 &  & 0.018 &  & 0.019 &  & 0.026 &  & 0.027 &  & 0.001 \\
$\mathrm{Na}_2$ & 0.016 &  & 0.015 &  & 0.013 &  & 0.012 &  & 0.013 &  & 0.012 &  & 0.008 &  & 0.008 &  & 0.000 \\
$\mathrm{O}_2$  & 0.041 &  & 0.041 &  & 0.031 &  & 0.031 &  & 0.006 &  & 0.007 &  & 0.028 &  & 0.029 &  & 0.001 \\
$\mathrm{S}_2$  & 0.049 &  & 0.049 &  & 0.037 &  & 0.037 &  & 0.007 &  & 0.007 &  & 0.036 &  & 0.037 &  & 0.000 \\
                &       &  &       &  &       &  &       &  &       &  &       &  &       &  &       &  &       \\
MEAN            & 0.042 &  & 0.042 &  & 0.029 &  & 0.029 &  & 0.011 &  & 0.011 &  & 0.024 &  & 0.025 &  & 0.000 \\ \bottomrule
\end{tabular}
\end{table*}

The $\Delta$ defined by eqn.~\ref{eq:delta} quantifies the difference of EOS curves in equilibrium volume and energy response with respect to the volume perturbation. The $\Delta$ values between PW and NAOs for 26 bulk systems are listed (Table~\ref{tab:crystal-delta}).
\begin{table}[]
\centering
\caption{Delta values with respect to the PW calculated by NAOs for 26 bulk systems (in meV/atom).}
\label{tab:crystal-delta}
\begin{tabular}{@{}lllll@{}}
\toprule
Bulk & 
\multicolumn{1}{c}{pVDZ} & 
\multicolumn{1}{c}{pVTZ$^-$} & 
\multicolumn{1}{c}{pVTZ} & 
\multicolumn{1}{c}{pVQZ$^=$} \\ \midrule
$\mathrm{AlAs}$ & 1.261 & 0.551 & 0.027 & 0.718 \\
$\mathrm{AlP}$  & 1.240 & 0.521 & 0.172 & 0.863 \\
$\mathrm{AlSb}$ & 0.848 & 0.597 & 0.180 & 0.325 \\
$\mathrm{AlN}$  & 3.459 & 0.873 & 0.125 & 1.722 \\
$\mathrm{BN}$   & 0.978 & 0.315 & 0.161 & 0.300 \\
$\mathrm{BP}$   & 2.529 & 0.810 & 0.017 & 0.596 \\
C               & 0.078 & 0.585 & 0.112 & 0.227 \\
$\mathrm{CdTe}$ & 0.899 & 0.533 & 0.215 & 0.519 \\
$\mathrm{CdS}$  & 0.500 & 0.939 & 0.101 & 0.720 \\
$\mathrm{CdSe}$ & 0.615 & 1.174 & 0.048 & 0.936 \\
$\mathrm{GaAs}$ & 1.348 & 0.573 & 0.173 & 0.520 \\
$\mathrm{GaP}$  & 1.141 & 0.781 & 0.134 & 0.686 \\
$\mathrm{GaSb}$ & 1.057 & 0.311 & 0.053 & 0.422 \\
$\mathrm{GaN}$  & 1.530 & 0.664 & 0.192 & 0.613 \\
$\mathrm{InP}$  & 1.194 & 0.837 & 0.204 & 0.832 \\
$\mathrm{LiF}$  & 0.585 & 1.166 & 1.227 & 0.126 \\
$\mathrm{MgO}$  & 2.625 & 0.412 & 0.177 & 0.413 \\
$\mathrm{MgS}$  & 0.593 & 1.070 & 0.121 & 0.683 \\
$\mathrm{NaCl}$ & 0.026 & 0.236 & 0.086 & 0.250 \\
Si              & 1.404 & 0.655 & 0.089 & 0.744 \\
$\mathrm{SiC}$  & 3.764 & 3.911 & 0.577 & 2.280 \\
$\mathrm{ZnSe}$ & 2.021 & 1.811 & 0.884 & 1.412 \\
$\mathrm{ZnTe}$ & 2.005 & 1.022 & 0.743 & 0.864 \\
$\mathrm{ZnO}$  & 5.058 & 2.381 & 2.084 & 1.306 \\
$\mathrm{ZnO}$w & 5.135 & 2.508 & 2.217 & 1.353 \\
$\mathrm{ZnS}$  & 2.530 & 2.319 & 1.245 & 1.673 \\
                &       &       &       &       \\
MEAN            & 1.709 & 1.060 & 0.437 & 0.812 \\ \bottomrule
\end{tabular}
\end{table}

The $\eta$-metrics measuring the band structure differences between PW and NAOs of 26 bulk systems are listed in Table~\ref{tab:crystal-eta}.
\begin{table}[]
\centering
\caption{Occupation weighted band structure differences with PW basis ($\eta$-metrics) calculated on 26 bulk systems (in eV)}
\label{tab:crystal-eta}
\begin{tabular}{@{}lllll@{}}
\toprule
Bulk & 
\multicolumn{1}{c}{pVDZ} & 
\multicolumn{1}{c}{pVTZ$^-$} & 
\multicolumn{1}{c}{pVTZ} & 
\multicolumn{1}{c}{pVQZ$^=$} \\ \midrule
$\mathrm{AlAs}$ & 0.017 & 0.008 & 0.004 & 0.006 \\
$\mathrm{AlN}$  & 0.065 & 0.040 & 0.022 & 0.024 \\
$\mathrm{AlP}$  & 0.020 & 0.013 & 0.008 & 0.008 \\
$\mathrm{AlSb}$ & 0.007 & 0.003 & 0.002 & 0.002 \\
$\mathrm{BN}$   & 0.004 & 0.002 & 0.001 & 0.002 \\
$\mathrm{BP}$   & 0.005 & 0.003 & 0.001 & 0.002 \\
C               & 0.004 & 0.003 & 0.002 & 0.002 \\
$\mathrm{CdS}$  & 0.021 & 0.006 & 0.009 & 0.003 \\
$\mathrm{CdSe}$ & 0.025 & 0.005 & 0.009 & 0.003 \\
$\mathrm{CdTe}$ & 0.048 & 0.018 & 0.018 & 0.008 \\
$\mathrm{GaAs}$ & 0.024 & 0.005 & 0.005 & 0.005 \\
$\mathrm{GaN}$  & 0.028 & 0.008 & 0.004 & 0.006 \\
$\mathrm{GaP}$  & 0.024 & 0.006 & 0.006 & 0.005 \\
$\mathrm{GaSb}$ & 0.012 & 0.003 & 0.003 & 0.003 \\
$\mathrm{InP}$  & 0.013 & 0.005 & 0.005 & 0.004 \\
$\mathrm{LiF}$  & 0.098 & 0.057 & 0.049 & 0.015 \\
$\mathrm{MgO}$  & 0.048 & 0.033 & 0.019 & 0.022 \\
$\mathrm{MgS}$  & 0.061 & 0.039 & 0.018 & 0.025 \\
$\mathrm{NaCl}$ & 0.053 & 0.038 & 0.034 & 0.017 \\
Si              & 0.005 & 0.003 & 0.001 & 0.003 \\
$\mathrm{SiC}$  & 0.011 & 0.008 & 0.002 & 0.006 \\
$\mathrm{ZnO}$  & 0.023 & 0.016 & 0.016 & 0.010 \\
$\mathrm{ZnO}$w & 0.024 & 0.016 & 0.016 & 0.011 \\
$\mathrm{ZnS}$  & 0.019 & 0.011 & 0.014 & 0.007 \\
$\mathrm{ZnSe}$ & 0.021 & 0.007 & 0.012 & 0.005 \\
$\mathrm{ZnTe}$ & 0.030 & 0.014 & 0.014 & 0.009 \\
                &       &       &       &       \\
MEAN            & 0.027 & 0.014 & 0.011 & 0.008 \\ \bottomrule
\end{tabular}
\end{table}

By manually lifting the Fermi level ($\varepsilon^A_\mathrm{f}$ and $\varepsilon^B_\mathrm{f}$) by 10 eV in eqn.~\ref{eqn:eta}, $\eta$-metrics measure the band structure in the range where more conduction bands are included. The data of $\eta_{10}$ benchmark for 26 bulk systems are listed in Table~\ref{tab:crystal-eta10}.
\begin{table*}[]
\centering
\caption{Occupation weighted band structure differences with PW basis ($\eta_{10}$-metrics) calculated on 24 bulk systems (in eV, Fermi levels are up-shifted for 10 eV)}
\label{tab:crystal-eta10}
\begin{tabular}{@{}lllllll@{}}
\toprule
Bulk &
  \multicolumn{1}{c}{pVDZ} &
  \multicolumn{1}{c}{pVTZ$^-$} &
  \multicolumn{1}{c}{pVTZ} &
  \multicolumn{1}{c}{pVTZ-2V} &
  \multicolumn{1}{c}{pVTZ-3V} &
  \multicolumn{1}{c}{pVQZ$^=$} \\ \midrule
$\mathrm{AlAs}$ & 0.228 & 0.068 & 0.052 & 0.047 & 0.034 & 0.042 \\
$\mathrm{AlN}$  & 0.106 & 0.050 & 0.023 & 0.020 & 0.027 & 0.027 \\
$\mathrm{AlP}$  & 0.207 & 0.067 & 0.063 & 0.033 & 0.031 & 0.030 \\
$\mathrm{AlSb}$ & 0.286 & 0.077 & 0.073 & 0.027 & 0.044 & 0.027 \\
$\mathrm{BN}$   & 0.114 & 0.016 & 0.002 & 0.003 & 0.003 & 0.006 \\
$\mathrm{BP}$   & 0.149 & 0.040 & 0.016 & 0.006 & 0.007 & 0.027 \\
C               & 0.102 & 0.015 & 0.023 & 0.005 & 0.006 & 0.014 \\
$\mathrm{CdS}$  & 0.575 & 0.065 & 0.056 & 0.035 & 0.033 & 0.019 \\
$\mathrm{CdSe}$ & 0.521 & 0.048 & 0.045 & 0.044 & 0.043 & 0.021 \\
$\mathrm{CdTe}$ & 0.446 & 0.062 & 0.051 & 0.050 & 0.039 & 0.017 \\
$\mathrm{GaAs}$ & 0.202 & 0.049 & 0.040 & 0.041 & 0.036 & 0.027 \\
$\mathrm{GaN}$  & 0.079 & 0.013 & 0.007 & 0.016 & 0.025 & 0.008 \\
$\mathrm{GaP}$  & 0.270 & 0.051 & 0.028 & 0.037 & 0.036 & 0.020 \\
$\mathrm{GaSb}$ & 0.359 & 0.047 & 0.044 & 0.028 & 0.034 & 0.018 \\
$\mathrm{InP}$  & 0.286 & 0.129 & 0.116 & 0.040 & 0.032 & 0.030 \\
$\mathrm{LiF}$  & 0.098 & 0.066 & 0.056 & 0.051 & 0.042 & 0.016 \\
$\mathrm{MgO}$  & 0.101 & 0.055 & 0.021 & 0.022 & 0.043 & 0.033 \\
$\mathrm{MgS}$  & 0.477 & 0.067 & 0.028 & 0.028 & 0.051 & 0.045 \\
$\mathrm{NaCl}$ & 0.181 & 0.130 & 0.294 & 0.141 & 0.133 & 0.029 \\
Si              & 0.331 & 0.087 & 0.064 & 0.046 & 0.028 & 0.048 \\
$\mathrm{SiC}$  & 0.159 & 0.037 & 0.017 & 0.016 & 0.008 & 0.019 \\
$\mathrm{ZnO}$  & 0.086 & 0.021 & 0.022 & 0.025 & 0.019 & 0.013 \\
$\mathrm{ZnO}$w & 0.071 & 0.022 & 0.021 & 0.023 & 0.018 & 0.013 \\
$\mathrm{ZnS}$  & 0.293 & 0.035 & 0.028 & 0.037 & 0.036 & 0.021 \\
$\mathrm{ZnSe}$ & 0.238 & 0.038 & 0.028 & 0.036 & 0.034 & 0.022 \\
$\mathrm{ZnTe}$ & 0.265 & 0.034 & 0.035 & 0.042 & 0.034 & 0.019 \\
                &       &       &       &       &       &       \\
MEAN            & 0.240 & 0.053 & 0.048 & 0.035 & 0.034 & 0.024 \\ \bottomrule
\end{tabular}
\end{table*}

\bibliographystyle{apsrev4-2}
\bibliography{refs}

\end{document}